\documentclass[10pt]{article}

           \usepackage{epsf}
           \usepackage{amsmath}
           \usepackage{amsfonts}
           \usepackage{amssymb}

\begin{document}

\title{String/M-branes for Relativists}

\author{Donald Marolf}

\maketitle

\centerline{Department of Physics, Syracuse University,\\
Syracuse, NY 13244, USA,}

\vspace{1cm}

\centerline{and}

\vspace{1cm}

\centerline{Institute for Theoretical Physics,}
\centerline{University of California, Santa Barbara, California, 93106, USA.}

\begin{abstract}
These notes present an introduction to branes in ten and eleven dimensional
supergravity and string/M-theory which is geared to an audience of traditional
relativists, especially graduate students and others with little background
in supergravity.  They are designed as a tutorial and not as a thorough
review of the subject; as a result, many topics of current interest are
not addressed.  However, a guide to further reading is included.
The presentation begins with eleven dimensional
supergravity, stressing its relation to 3+1 Einstein-Maxwell theory.
The notion of Kaluza-Klein compactification is then introduced, and is
used to relate the eleven dimensional discussion to supergravity in
9+1 dimensions and to string theory.  The focus is on type IIA
supergravity, but the type IIB theory is also addressed, as is the
T-duality symmetry that relates them.  Branes in both 10+1 and 9+1 dimensions
are included.  Finally, although the details are not discussed, 
a few comments are provided on the relation between supergravity and
string perturbation theory and on black hole entropy.  
The goal is to provide traditional
relativists with a kernel of knowledge from which to grow their
understanding of branes and strings.
\end{abstract}

\vfil \eject

\vspace*{-1cm}
\centerline{\bf TABLE OF CONTENTS}
\bigskip
\contentsline{section} {\numberline {1.} Introduction}{3}
\bigskip
\contentsline{section} {\numberline {2.} 
Supergravity in Eleven Dimensions}{4}
\medskip
\contentsline{subsection} {\numberline {2.1.} On
$n$-form gauge fields}{4}
\smallskip
\contentsline{subsection} {\numberline {2.2.} Dynamics
}{7}
\smallskip
\contentsline{subsection} {\numberline {2.3.} 
Supersymmetry and BPS States}{9}
\bigskip
\contentsline{section} {\numberline {3.} 
Supergravity in Eleven Dimensions
M-branes: The BPS Solutions}{12}
\medskip
\contentsline{subsection} {\numberline {3.1.} 
The 3+1 Majumdar-Papapetrou solutions}{13}
\smallskip
\contentsline{subsection} {\numberline {3.2.} 
Brane solutions in eleven dimensions}{15}
\smallskip
\contentsline{subsection} {\numberline {3.3.} 
Brane Engineering}{20}
\bigskip
\contentsline{section} {\numberline {4.}
Kaluza-Klein and Dimensional Reduction}{23}
\medskip
\contentsline{subsection} {\numberline {4.1.} 
Some remarks on Kaluza-Klein reduction}{24}
\smallskip
\contentsline{subsection} {\numberline {4.2.} 
Kaluza-Klein in (super)gravity}{24}
\smallskip
\contentsline{subsection} {\numberline {4.3.} 
On 9+1 Dynamics: Here comes the dilaton}{27}
\bigskip
\contentsline{section} {\numberline {5.} 
Branes in 9+1 type II Supergravity}{30}
\medskip
\contentsline{subsection} {\numberline {5.1.} 
The type IIA branes}{30}
\smallskip
\contentsline{subsection} {\numberline {5.2.} 
On brane singularities}{33}
\smallskip
\contentsline{subsection} {\numberline {5.3.} 
T-duality and the type IIB theory}{35}
\bigskip
\contentsline{section} {\numberline {6.}
Some Remarks on D-brane Perturbation Theory}{37}
\medskip
\contentsline{subsection} {\numberline {6.1}
Background field expansions and perturbative string theory}{37}
\smallskip
\contentsline{subsection} {\numberline {6.2.} 
Strings and D-branes}{40}
\smallskip
\contentsline{subsection} {\numberline {6.3.} 
On branes and perturbative expansions}{41}
\smallskip
\contentsline{subsection} {\numberline {6.4.} 
A few words on black hole entropy}{43}
\bigskip
\contentsline{section} {\numberline {Acknowledgements}
}{44}
\medskip
\contentsline{section} {\numberline {Appendices}
}{}
\medskip
\contentsline{subsection} {\numberline {
A.} Guide to Further Reading}{45}
\smallskip
\contentsline{subsection} {\numberline {
B.} Conventions} {46}
\bigskip
\contentsline{section} {\numberline {References} 
}{46}

\vfil \eject

\section{Introduction}
\label{intro}
This contribution to the Proceedings of the 3rd Mexican School
on Gravitation and Mathematical Physics (Mazatl\'an, Mexico) is intended to be 
an introduction to the branes which have
contributed so much to studies of string/M-theory in
recent years.  The goal of this work is to open string/M-theory
to traditional relativists and students of 
general relativity by describing a few features in an accessible language.
Our aim here is not in any way to be complete.
In fact, we will purposely leave out many of the points which practicing
string theorists deem to be most relevant, such as the detailed
setup of D-brane perturbation theory and its relation to super
Yang-Mills theory, S-duality, 
and the Maldacena conjecture.  Some general commentary
on the conceptual framework within which D-brane perturbation theory is
to be viewed will, however, be presented in section \ref{pert}.
The hope is that this will provide a chunk of material that
a member of the relativity community can latch on to and use to develop
a perspective on both branes in particular, and string/M-theory in
general.   From here, one can move on to more advanced topics.  A guide
to further reading is given in appendix \ref{read}.        

We will focus here on aspects of branes which can be described in terms of
supergravity.  To provide a context for this discussion, 
the first part of the text will introduce 
various aspects of supergravity. 
We draw heavily on Polchinski's
treatment \cite{Joe}, though the style is (hopefully)
more adapted to the current audience.   We begin in
section \ref{11sugrav} with a discussion of 
supergravity in eleven dimensions, which is both a particularly relevant
and a particularly simple case.  We then introduce the
associated branes in section \ref{Mbr}.

The case of ten dimensional supergravity is also central to our
mission.  We choose to approach this subject via Kaluza-Klein reduction of
eleven dimensional supergravity.  Section \ref{KKred} thus begins with
some introductory remarks on Kaluza-Klein compactifications 
and then constructs a supergravity theory (the type IIA theory) 
in 9+1 dimensions.  Section \ref{10branes} introduces the branes
of type IIA theory and then comments briefly on type IIB supergravity and
the so-called T-duality symmetry that relates the two
theories.  We close with some remarks on D-brane
perturbation theory and a few words about black hole
entropy via D-branes in section \ref{pert}.

In broad outline, our discussion will follow the lectures originally
given in Mazatl\'an.  For the sake of both the author and the reader, no
attempt will be made to be substantially more comprehensive or to include many
relevant points which are not truly central to our discussion.
However, the details of the presentation will be rather different and, 
in particular, the connections with eleven dimensional supergravity
will be emphasized far more heavily than in the original lectures. 
This will in fact provide a more comfortable perspective for relativists, 
as the various classical brane solutions
are somewhat less singular in eleven dimensional
supergravity.  As a result, the weight of the discussion has been shifted
relative to the original lectures
away from type IIB supergravity and toward the
IIA case which is more directly related to the eleven dimensional
theory.

The literature on branes and string theory is rather
vast.  Our goal here is to provide a tutorial and not a complete
review.  As a result, our referencing of the original works
will at times be rather spotty.  For more complete reference
lists, see the reviews mentioned in appendix \ref{read}, especially
\cite{YR}.

\section{Supergravity in Eleven Dimensions}
\label{11sugrav}

Before diving into the details, a few words of orientation are
in order.  We will shortly see that supergravity in
eleven (10+1) dimensions is really not much more complicated than the 3+1
Einstein-Maxwell theory of Einstein-Hilbert gravity coupled to
Maxwell electrodynamics.   The same is not 
as true of supergravity in lower dimensions.  In ten  (9+1)
dimensions and below, many interesting supergravity theories
contain a so-called dilaton field which couples non-minimally to
the Maxwell-like gauge fields.  As a result, the equivalence principle
does not hold in such theories and different fields couple to
metrics that differ by a conformal factor.
However, in eleven dimensions, properties of the supersymmetry
algebra guarantee that any supergravity theory
containing no fields
with spin higher than two\footnote{Except
for anti-symmetric tensor fields, which propagate on curved manifolds
without the constraints associated with other higher spin fields.}, has
no dilaton.  In fact, there is a a unique supergravity theory in
eleven dimensions and it contains only three fields:  the metric,
a U(1) (i.e., abelian, Maxwell-like) gauge field, and a spin 3/2
gravitino.  

\subsection{On $n$-form gauge fields}
\label{forms}

We first address just the bosonic part of eleven-dimensional supergravity, 
setting the fermionic fields to zero.  The differences between
this truncated
theory and 3+1 Einstein-Maxwell theory amount to just differing numbers
of dimensions.  This happens in two ways:  The first is
the obvious fact that
the theory lives in a 10+1 spacetime instead of a 3+1 spacetime.  The 
second is that the gauge field is slightly `larger' than that of Maxwell
theory.  Instead of having a {\it vector} (or, equivalently, a one-form)
potential, the potential is a 3-form: $A_3 = \frac{1}{3!}
A_{ 3, \alpha \beta \gamma}
dx^\alpha \wedge dx^\beta \wedge dx^\gamma$.  See appendix \ref{conv}
for a discussion of the conventions used here for $n$-forms.  

We will be seeing a lot of $n$-form potentials below.  Although they
may at first seem unfamiliar, they are in fact a very natural
(and very slight) generalization of Maxwell fields.  An $n$-form
gauge potential $A_n$ is associated with an $(n+1)$-form
field strength of the form $F_{n+1} = dA_n$, where $d$ is the exterior
derivative.  As a result, the field strength satisfies a Bianchi 
identity $d F_{n+1} =0$.   As with the familiar Maxwell field, there is 
an associated set of gauge transformations
\begin{equation}
A_n \rightarrow A_n + d \Lambda_{n-1}
\end{equation}
where $\Lambda_{n-1}$ is an arbitrary $(n-1)$ form.   Such
gauge transformations leave the field strength $F_{n+1}$ invariant.

In $D$ spacetime dimensions (generally taken to be $11$ in this section), the
equation of motion for such a gauge field is typically of the
form 
\begin{equation}
\label{current}
d\star F_{D-(n+1)} = J_{D-n}, 
\end{equation}
where, in a slight abuse of notation, 
$\star F_{D-(n+1)}$ denotes the $D-(n+1)$ form that is the hodge-dual of
$F_{(n+1)}$ and
$J_{D-n}$ is a $(D-n)$-form current that serves as a source for the field
strength.  Thus, the natural coupling between an $n$-form
gauge field and its current is of the form $\int_M A_n
\wedge J_{D-n}$, where $M$
denotes the spacetime manifold.  

As usual, gauge symmetry implies that the current is
conserved.  However, current conservation for a $(D-n)$-form current with
$D-n>1$ is, in a certain sense, a much stronger statement than 
conservation of the current in 3+1 Maxwell theory.  
Note that the analogue of Gauss'
law in the present context is to define the charge $Q_B$ contained in
a $(D-n)$-ball $B$ by the integral
$Q_{D-n} = \int_{\partial B} \star F_{D-(n+1)}$ over the boundary $\partial B$
of that ball.  Now, suppose that the current $J_{D-n}$ in fact
vanishes in a neighborhood of the surface $\partial B$.  Then by stokes
theorem and equation (\ref{current}) we can deform the surface $\partial B$ in
any way we like and, as long as the surface does not encounter any
current, the total charge $Q_{D-n}$ does not change.

Now, in
familiar 3+1 Maxwell theory, charge is measured by integrals over
2-surfaces.  This is associated with the fact that
an electrically charged particle sweeps out a worldline in spacetime. 
Note that any sphere which can be collapsed
to a point without encountering the worldline of the particle must
enclose zero net charge.  The important fact is that,
in four dimensions, there are two-spheres which `link' with 
any curve and which cannot in fact be shrunk to a point without encountering
the particle's worldline.  
In contrast, circles do not link with worldlines in
3+1 dimensions.  For this reason, particles in 3+1 dimensions cannot be
charged under any gauge field whose field strength is, for example, a 
3-form.  In considering a one-form field strength, note that
 3-spheres generically {\it intersect} with worldlines
and so do not {\it enclose} charge as 2-spheres do.     
This illustrates a general relation between a gauge field
and the associated charges: 
unless the world-volume of an object can link with
surfaces of dimension $D-(n+1)$, it cannot be electrically charged under
an $n$-form gauge potential $A_n$.  

While we are here, we may as well work out this counting.  Let us
suppose that we have an $n$-form gauge field $A_n$ in $D$ spacetime
dimensions.  Then, we must integrate $\star F_{D-(n+1)}$ over a $D-(n+1)$
surface in order to calculate the charge.  Now, in $D$ dimensions, 
surfaces of dimensions $k$ and $m$ can link if $k+m +1 =D$ (i.e., 
curves and curves in three dimensions, 2-surfaces and worldlines in
four dimensions, etc.).  Thus, non-zero electric charge of $A_n$ is
associated with $n$ dimensional worldvolumes.  Such objects are generically
known as `$p$-branes' (as higher dimensional generalizations
of the term membrane). Here, $p$ is a the number of spatial
dimensions of the object; i.e., the electric charge of
an $n$-form gauge potential is carried by $(n-1)$-branes, whose
world volume has $n-1$ spatial dimensions and time.  This is how
strings, membranes, and other branes will arise in our discussion
of supergravity.

Note that, although the $p$-branes are extended objects, the concept
of a charge {\it density} on the branes is not appropriate.  Recall
that the charge is measured by any $D-(p+2)$ surface surrounding the
brane and that, by the above charge conservation argument, the
charge detected by any such surface must necessarily be the same.
Thus, the equations of motion tell us that moving the $D-(p+2)$
surface {\it along} the brane cannot {\it ever} change the flux through
the surface;  see Fig. 1 below.  
Thus, `non-uniform' $p$-branes cannot exist! 
The proper concept here is to assign to such a $p$-brane only
one number, the total charge.  It simply happens that the particular
type of charge being measured is somewhat less local than
the familiar electric charge; it is fundamentally associated with
$p+1$ dimensional hypersurfaces in the spacetime.

\vbox{
\centerline{\epsfbox{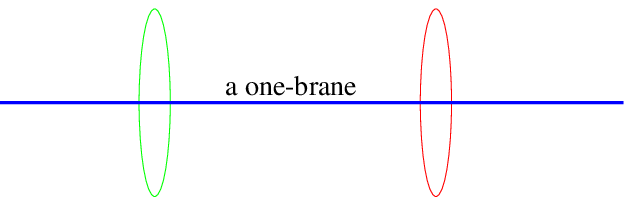}}
\centerline{Fig. 1. By charge conservation, both circles necessarily
capture the same flux.}}

\medskip

As a small complication, we will be interested not only in electric
charges, but also in magnetic charges.  Indeed, in supersymmetric
string theory, both electric and magnetic charges appear to be on
an equal footing.  A useful point of contact for the present discussion is
to realize that, in a certain sense, both electrically and magnetically
charged `objects' occur in pure Einstein-Maxwell theory without any
matter fields.  These are just the electrically and magnetically
charged eternal black hole solutions.  Although the Maxwell field
satisfies both $dF=0$ and $d\star F=0$ at every non-singular point of
such spacetimes, the black holes can still be said to `carry charge' due to
topological effects:  the electric or magnetic flux starts in one
asymptotic region, funnels through the Einstein-Rosen bridge at
the `throat' of the black hole, and out into the other asymptotic
region\footnote{As a result, the electric charge of the black hole measured in
one asymptotic region is the opposite of the charge measured in the 
other asymptotic region.  However, this need not trouble us so long
as we understand that we must first orient ourselves by
picking an asymptotic region in order to discuss  the notion of charge.}.
Note that black holes (i.e., point-like or zero-brane objects) may carry
both electric {\it and} magnetic charge for the Maxwell field in 3+1 dimensions.

The counting of dimensions
for magnetic charges proceeds much like the counting for electric
charges.  To define what we mean by a magnetic charge, we recall
that hodge duality $F \rightarrow \star F$ in Maxwell theory interchanges
electric and magnetic charge.  Thus, since electric charge is
associated with integrals of $\star F_{D-(n+1)}$, magnetic charge is defined
by integrating the field strength $F_{n+1}$ itself over an $n+1$
surface.  In $d$ dimensions, an $n+1$ surface can link with
$D-n -2$ worldvolumes, or $(D -n -3)$-branes.  As a check, for
3+1 Maxwell theory, we have magnetic $4- 1 - 3=0$ branes.

Let's take a look at the eleven-dimensional context.  Without knowing
anything more about supergravity than we already do, we can expect
two types of `objects' to be of particular interest from the point of
view of the 3-form gauge field $A_3$.  There may be 2+1 electrically
charged objects (2-branes) and $(D-n-3) +1 = (11-3-3)+1 = 5+1$ dimensional
magnetically charged objects (5-branes).  Since there are no
explicit charges in the theory, these `objects' (if they
exist) must be black-hole like `solitonic' solutions.  We will see
below that black two-brane and black five-brane solutions carrying
the proper charges do indeed exist in eleven-dimensional supergravity.
What is more, and what is different from lower dimensional
supergravity, is that the horizons of these black branes
remain smooth in the extremal limit
of maximal electric or magnetic charge.  The extremal versions of these
brane-solutions are what are usually referred to as `the M-theory
two-brane' or `M2-brane,'  and `the M-theory five-brane' or
`M5-brane.'  We will discuss these
in more detail in section \ref{branes}.

\subsection{Dynamics}

I hope the discussion of section \ref{forms} has provided
some orientation to supergravity in eleven dimensions.  Now, 
however, it is time to fill in a few details.  For example, 
it is appropriate to write down the full dynamics of the 
system.  This is conveniently summarized by the action \cite{CJS}
  
\begin{eqnarray}
\label{eleven}
S&=& \frac{1} {2 \kappa_{11}^2 } \int d^{11} x \  e \Bigl[
\bigl( R - \frac{1}{2} |F_4|^2 \bigr)  \cr
&-& \frac{1}{2^3 \cdot 4!} (\overline{\psi}_\alpha\Gamma^{\alpha \beta
\gamma \delta \sigma \lambda}
\psi_\lambda + 12 \overline{\psi}^\beta \Gamma^{\gamma \delta} \psi^\sigma)
(F + \hat F)_{\beta \gamma \delta \sigma} \cr
&-& \overline{\psi}_\alpha
\Gamma^{\alpha \beta \gamma} 
D_\beta\bigl(\frac{1}{2}(\omega + \hat \omega) \bigr)\psi_\gamma \Bigr]
- \frac{1}{12\kappa_{11}^2}
\int A_3 \wedge F_4 \wedge F_4.
\end{eqnarray}
Here $R$ is the Ricci scalar of the metric $g_{\alpha \beta}$, $e^a_\alpha$
is the vielbein which squares to $g_{\alpha \beta}$ (and $e$ is its
determinant), $A_3$
is the three-form field discussed in the previous section,
and $\psi$ is the spin 3/2 gravitino.  Here we use the notation
\begin{eqnarray}
\hat \omega_{\alpha ab} &=& \omega_{\alpha ab} + \frac{1}{8}
\overline{\psi}^\beta \Gamma_{\beta \alpha ab \gamma} \psi^\gamma, \cr  
\hat{F}_{\alpha \beta \gamma \delta} &=& F_{\alpha \beta \gamma \delta}
- 3 \overline{\psi}_{[\alpha} \Gamma_{\beta \gamma} \psi_{\alpha]}.
\end{eqnarray}
In the above, Greek letters ($\alpha, \beta,...$) denote spacetime indices
and Latin letters ($a,b,...$) denote internal indices.   The square
brackets $[...]$ indicates a completely antisymmetric sum over permutations
of the indices, divided by the number of terms.
Our conventions for spinors and
$\Gamma$-matrices are those of \cite{GSW}.  We will not state
them explicitly here as spinors only make appearances in this section and
section \ref{bps} and, in both cases, the details can be safely glossed
over.

This looks a little complicated, but let's
take a minute to sort through the various terms.
We'll begin with the least familiar part: the gravitino.
Since our attention here will be focused on classical solutions, we
will be able to largely ignore the gravitino.  The point
here is that the gravitino is a fermion and, due to
the Pauli exclusion principle, fermion fields do not have
semi-classical states of the same sort that bosonic fields do.
It is helpful here to think about the electron field as an
example.  There are, of course, states with a large number
of electrons that are well described by a classical charged
fluid.  However, because of the exclusion principle,
there are no semi-classical coherent states of the electron field 
itself; i.e., no
states for which the dynamics is well described by a
{\it classical spinor field}.  In the same way, we might
expect that there are states of the gravitino field that
are well described by some sort of classical fluid, but we should
only expect the classical action (\ref{eleven}) to be
a good description of the dynamics when the gravitino field
vanishes.  Thus, we will set $\psi =0$ throughout most of
our discussion.  This is self-consistent as setting $\psi =0$
in the initial data is enough to guarantee $\psi = 0$ for all time.

A study of (\ref{eleven}) shows that the dynamics of the solutions
for which $\psi =0$ can be obtained by simply setting $\psi$
to zero in the action.  This simplifies the situation
sufficiently that it is worth rewriting the action as:
  
\begin{equation}
\label{11B}
S_{\rm bosonic} = \frac{1} {2 \kappa_{11}^2 } \int d^{11} x (-g)^{1/2}
\bigl( R - \frac{1}{2} |F_4|^2 \bigr) - \frac{1}{6}
\int A_3 \wedge F_4 \wedge F_4. 
\end{equation}
This sort of presentation, giving only the bosonic terms, 
is quite common in the literature and
is sufficient for most solutions of interest\footnote{Typically, 
an interesting bosonic solution in fact corresponds to 
one of the quantum states in a supersymmetry multiplet.
The supersymmetry algebra can often be used to construct
from the original solution the related spacetimes in which
the fermions are excited.  See for example \cite{mult}.}.
Now that the gravitino has been set to zero, we see that
our action contains only three terms: the Einstein (scalar
curvature) term $R$, the Maxwell-like term $F_4^2$, and
the remaining so-called `Chern-Simons term.'  

The Chern-Simons
term has the same form ($A \wedge F \wedge F$) as the action for 2+1
Chern-Simons theory, but in the current eleven dimensional
context it is not a topological invariant.  Its variation is
not a total divergence and it contributes to the equations of motion.
Suppose that we couple a source to (\ref{11B}) and
vary the resulting action with respect to $A_3$.  The equation of motion
is of the form
\begin{equation}
d(\star F_4 + (const) F_4 \wedge F_4) = J.
\end{equation}
Thus, the charge conservation 
arguments of section \ref{forms} continue to hold, but with
the charge computed from integrals of $\star F_4 + F_4 \wedge F_4$ and not
just $\star F$.   This is quite important (see e.g. \cite{Town})
for certain features of brane physics, such
as some types of brane intersections and is essential for the BPS
bound to hold in eleven dimensions \cite{GHT}.  
Nevertheless, we will be able
to ignore the Chern-Simons term in our discussion below since we will consider
only relatively simple intersections.  Specifically
we note that since $dF =0$, the variation of the 
Chern-Simons term gives a contribution to the equations of motion
proportional to $F_4 \wedge F_4$. Thus, whenever there are 4 or more
linearly independent vectors $k$ at each point such that
$k^\alpha F_{\alpha \beta \gamma \delta} =0$, we have $F_4 \wedge F_4 =0$
and the Chern-Simons term does not affect the equations of motion.
For the cases we consider below, this property is
satisfied as all of the non-vanishing components of $F$ will
lie in a subspace of dimension seven or less.

\subsection{Supersymmetry and BPS States}
\label{bps}

Some comments are now in order on the subject of supersymmetry, so
that we may introduce (and then use!) the concept of BPS states.
Again, I would like to begin with a few heuristics to provide a rough
perspective for the traditional relativist.

We will
see that BPS solutions are closely related to extremal solutions;
in particular, to extremally charged solutions. 
As a result, most of our intuition from extreme Reissner-N\"ordstrom
solutions carries over to
the general BPS case.  A particularly useful property 
is that any BPS solution has a Killing vector which
is either timelike or null.  If, furthermore, the solution is
asymptotically flat in the usual strict sense\footnote{If the 
asymptotically flat region has the topology of ${\bf R^n}$
minus a compact set.} then in fact this Killing vector field
must be timelike \cite{GH}.

The setting for any discussion of BPS solutions is 
the class of supergravity solutions with a certain amount of asymptotic
flatness, though we will not go into the details of the boundary
conditions and fall-off rates here.  For the remainder of this
work, we will follow the usual terminology of string theory and
use a somewhat less restrictive notion of
asymptotic flatness than is common in traditional relativity.
The main difference is that we will {\it not} require the 
topology of the asymptotic region to be of the form $R^n$
minus a compact set.  Instead, we will allow it to be of the
form $(R^n - \Sigma) \times Y$, for any compact set $\Sigma
\subset R^n$, $Y$ any homogeneous manifold, and\footnote{
We impose this last condition as a substitute for spelling out the
fall-off conditions.  As is well known (see e.g. \cite{2+1}), 
the asymptotic fall-off conditions
for 2+1 (and smaller) dimensional spacetimes are qualitatively
different from those in higher dimensions.}   $n \ge 4$.
We will use the term `strict asymptotic flatness' to indicate the
special case where the homogeneous manifold $Y$ is a single point.

In the setting of pure gravity, one would expect that (under
appropriate fall-off conditions at infinity) such spacetimes
would exhibit asymptotic symmetries that correspond to the 
Poincar\'e group in the appropriate number of
spacetime dimensions together with the symmetries of $Y$.  Some particular
solutions in this class will even have Killing vectors which make 
some subgroup of the Poincar\'e group into an {\it exact} symmetry of the
spacetime; e.g. the rotation subgroup in spherically symmetric cases.
Poincar\'e group.  These are just the metric and matter fields that are
invariant under some non-trivial continuous subgroup of the diffeomorphism
group.

Now, supersymmetry is best thought of as an (anti-commuting)
extension of the diffeomorphism group.  Indeed, diffeomorphisms
form a subgroup of the supersymmetry gauge transformations and, 
in the asymptotically flat setting just described, the
asymptotic Poincar\'e transformations will be a subgroup of
the asymptotic supersymmetry transformations.  Solutions that
are invariant under a subgroup of the supersymmetry transformations
containing non-trivial anti-commuting elements are said to have
a `Killing Spinor' and are known as BPS (Bogomuln'yi-Prasad-Sommerfeld)
solutions.   We will see that they are closely related to extremal
solutions.

Having oriented ourselves with this intuitive introduction, it 
is now time to examine the details of the eleven-dimensional
supersymmetry transformations and their algebra.  The infinitesimal
supersymmetry transformations are in one-to-one correspondence with
Grassmann valued (Majorana\footnote{i.e., satisfying the reality
condition $\eta^* = B\eta$
where $B = \Gamma^3 \Gamma^5 ... \Gamma^9$ and $*$ denotes complex
conjugation.}) spinor fields $\eta(x)$.  The transformation
associated with $\eta$ is given by

\begin{eqnarray}
\label{susy}
\delta e^a_\alpha &=& \frac{1}{2} \overline \eta \Gamma^a \psi_\alpha, \cr
\delta A_{\alpha \beta \gamma} &=&  - \frac{3}{2} 
\overline \eta \Gamma_{[\alpha \beta }\psi_{\gamma]},\cr
\delta \psi_\alpha &=&  D_\alpha(\hat{\omega}) \eta
+ \frac{\sqrt{2}}{(4!)^2}\bigl( \Gamma_M^{PQRS} - 8 \delta_M^P \Gamma^{QRS}
\bigr) \eta \hat{F}_{PQRS} \equiv \hat D_\alpha \eta,
\end{eqnarray}
where the last line defines the supercovariant derivative $\hat D_\alpha$
acting on the spinor $\eta$.

The details of the  supersymmetry
transformations are not particularly important
for our purposes.
What {\it is} important is the general structure.  Note that
the variation of the vielbein $e$ involves the gravitino $\psi$, but
then the variation of the gravitino involves the connection $\hat \omega$
which contains derivatives of the vielbein.  Similarly taking two
variations of the gauge field $A_3$, we find terms involving derivatives
of the gauge field.  It turns out that, in fact, the proper second
variations give just diffeomorphisms of the spacetime. 

Recalling that the variation of $A_3$ contains $\psi$, we 
also note that the first variation of the gravitino field
involves a derivative of the spinor $\eta$.  Thus, the second variation
of $A_3$ is something that involves the derivative of some gauge parameter.
With the proper choice of spinors $\eta$, one can construct a second
supersymmetry variation that gives just the usual gauge transformation
$A_3 \rightarrow A_3 + d \Lambda_2$ on the gauge field.  Thus, 
both diffeomorphisms and gauge transformations are in fact contained
in the spacetime supersymmetry algebra.  The supersymmetry algebra can
be thought of as a sort of `square root' of the diffeomorphism and gauge
algebras.   The fact that diffeomorphisms and gauge transformations
are expressed as squares leads to extremely useful positivity
properties.

We will not need the details of the local supersymmetry algebra below.
However, it is useful to display the 
algebra of the asymptotic supercharges. 
Just as for the diffeomorphisms and
gauge transformations, the asymptotic supersymmetries  
lead, in the asymptotically flat context, 
to conserved `supercharges.'  In fact, for the eleven dimensional
case, there are several relevant notions of the asymptotic algebra.
This is because there are interesting $p$-branes with several
values of $p$.  Thus, there are several interesting classes of 
`asymptotically flat' structures associated with different choices
of the homogeneous manifold $Y = {\bf R}^{11-p}$ in our
generalization of strict asymptotic flatness.

However, all of these algebras are rather similar.
If $Q$ is the generator of supersymmetry transformations,
so that the asymptotic versions of the
transformations above are generated by taking super Poisson
brackets with $Q \overline \eta$, then the algebra associated with
the $p$-brane case has the general form

\begin{equation}
\{Q_a^A,\overline{Q}^{Bb}\}_+ = -2 P_\mu \Gamma^{\mu b}_a \delta^{AB}
 - 2 i Z^{AB} \delta_a^b, 
\end{equation}
where we have used $a,b$ for the internal spinor indices.
Here, $P_\mu$ are the momenta per unit $p$-volume and $Z^{AB}$
is an antisymmetric real matrix associated with the asymptotic
gauge transformations.  In
particular, the eigenvalues of $Z$ are of the form $\pm i q$
where $q$ is the appropriately normalized
charge carried by the $p$-brane.  Our notation
reflects that fact that it is natural to split the SUSY generator Q, 
which is an eleven dimensional Majorana fermion, into a set
of $(11-p)$ dimensional fermions $Q^A$.   Thus, the indices $a,b$
take values appropriate to spinors in $(11-d)$ dimensions.

The most important property of this algebra is that it implies
the so-called BPS bound on masses and charges.  To get an idea of
how this arises,
recall that while $\overline {Q} Q$ is a Lorentz invariant, it is
$Q^\dagger Q$  that is a positive definite operator.  Thus, a positivity
condition should follow by writing the algebra in terms of $Q^\dagger$ and
$Q$.  For simplicity, let us also choose an asymptotic
Lorentz frame such that energy-momentum of the spacetime is aligned with
the time direction: $P_\mu = T \delta_{\mu0}$, where $T$ is the
brane tensions, or mass per unit $p$-volume.  The algebra then
takes the form

\begin{eqnarray}
\label{simp}
\{Q_a^A,Q^{\dagger Bb}\}_+ &=& 2T \delta^{AB} \delta_a^b + 2i Z^{AB}
\Gamma^{0b}_a
\end{eqnarray}

It is useful to adopt the notation of quantum mechanics, even though
we are considering classical spacetimes.  Thus, we describe a spacetime
by a state $|\psi\rangle$ and we let the generators $Q$ act on that
state as $Q|\psi\rangle$.   Contracting the above relation (\ref{simp})
with $\eta_{bB}$ and $\eta^{\dagger a}_A$ for a set of spinor
fields $\eta_A$, taking
the expectation value in any state, and using the positivity of the
inner product and the fact that the eigenvalues of $\Gamma^{0b}_a$ are
$\pm1$ yields the relation
\begin{equation}
\label{BPS}
T \ge  |q|.
\end{equation}
See \cite{GHT} for a full classical supergravity gravity 
derivation in the context of magnetic charge in eleven dimensions and
\cite{GH} for a complete derivation in classical
$N=2$ supergravity in four dimensions.  See also \cite{alg} 
for details of the above argument in the four dimensional context.

This is the BPS bound.  A spacetime in which this
bound is saturated is called a BPS spacetime and the corresponding
quantum states are known as BPS states.  Note that, from our above
argument, a state is BPS only if it is annihilated by one of the
supersymmetry generators; that is, if the spacetime is invariant under
the transformation (\ref{susy}) for some spinor $\eta$.  
The converse is also true; any asymptotically flat
spacetime which is invariant under
some nontrivial supersymmetry transformation is BPS.
Given a solution $s$ and a spinor $\eta$
for which the transformation (\ref{susy}) vanishes on $s$, one says
that $\eta$ is a {\it Killing spinor} of $s$.  
Since the gravitino $\psi$ vanishes for a bosonic solution, in this
context we see from (\ref{susy}) that $\eta$ is
a Killing spinor whenever it is supercovariantly constant; i.e when it 
satisfies $\hat D_\alpha \eta = 0$.

Consideration of the full local spacetime supersymmetry algebra in a BPS
spacetime with Killing spinor $\eta$ shows \cite{GH}
that not only is the energy determined by the charge, but
that the spacetime in fact has a non-spacelike Killing field 
$\overline{\eta} \Gamma^\mu \eta$.
When the spacetime is asymptotically flat in all directions
(i.e., the fields decay in an asymptotic region diffeomorphic to
${\bf R}^n$ minus some compact set), this Killing field is
in fact timelike. The structure of the argument is very much
like the Witten proof of the positive energy theorem \cite{Witten}. 

The bound (\ref{BPS}) is reminiscent of the extremality bound for
Reissner-N\"ordstrom black holes.  It turns out that the
relationship is a strong one.
Given the similarity of eleven dimensional supergravity to
Einstein-Maxwell theory, it will come as no surprise that
there is a supergravity theory in 3+1 dimensions that 
contains Einstein-Maxwell theory, together with a few extra fields.
When the extra fields vanish on an initial slice, they remain zero
for all time.  Thus,  
Einstein-Maxwell theory is a `consistent truncation' of
the supergravity.  In this context, the BPS bound and the
extremality bound for charge coincide when there is no angular
momentum.  Thus, any asymptotically
flat solution of Einstein-Maxwell theory with extremal charge and vanishing
angular momentum defines a BPS solution of the supergravity.

In general, any BPS black hole solution will be extremal, though
the converse is not always true.  An important example occurs
in four dimensions where all BPS states must have zero angular momentum.
Thus, the 3+1 extreme Kerr solution is not BPS.

Now that we have come to terms with supersymmetry, we can
proceed to ignore fermions completely in the sections below.

\section{M-branes: The BPS Solutions}
\label{Mbr}

Although we wish to focus on the eleven dimensional case, as
indicated above supersymmetry and supergravity can also be considered in less
than eleven dimensions; for example, in 3+1 dimensions.
In that case, any asymptotically flat solution of Einstein-Maxwell
theory with extremal charge and zero angular momentum is a BPS solutions
of 3+1 supergravity.  But this is just the class
of Majumdar-Papapetrou solutions, which
consist of some number of extremal Reissner-N\"ordstrom black
holes in static equilibrium.  Thus, the Majumdar-Papapetrou solutions
are a more familiar analogue of the eleven dimensional M-brane
solutions which we wish to discuss.  We therefore 
present a brief review of the Majumdar-Papapetrou solutions in
section \ref{MPsols}
below as an introduction to the world of M-branes.  We will examine the 
M-branes themselves in section \ref{branes} and find that they
strongly resemble the Majumdar-Papapetrou solutions.

\subsection{The 3+1 Majumdar-Papapetrou solutions}
\label{MPsols}

Recall that the Reissner-N\"ordstrom solution with
mass $M$ and charge $Q$
takes the form

\begin{equation}
\label{RNmetric}
ds^2 = - (1 - \frac{2MG}{R} + \frac{GQ^2}{R^2}) dt^2 +
\frac{1}{1 - \frac{2MG}{R} + \frac{GQ^2}{R^2}} dR^2 + R^2 d\Omega^2_2,
\end{equation}
with $R$ the usual Schwarzschild radial coordinate, $t$ the Killing
time, and $d\Omega^2_2$ the metric on the unit two-sphere.
Here, $Q$ and $M$ are the charge and mass of the black hole, with
$Q$ measured in units of $\sqrt{ (mass)(length)}$ as is natural in classical 
mechanics with $c=1$ but $G\neq 1$.  The factors of Newton's
constant $G$ have been left explicit for consistency
with the rest of this exposition.  The extremal situation
is $Q=M$ and, in this case, the solution is controlled
by a single length scale $r_0 =  GM = GQ$.  Since $r_0$ is defined by
the charge, we will
refer to it as the ``charge radius" of the black hole.
The metric simplifies to take the form

\begin{equation}
\label{XRNmetric}
ds^2 = - (1 -  {{r_0}/R})^2 dt^2 +
(1 - {{r_0}/R})^{-2} dR^2 + R^2 d\Omega^2_2.
\end{equation}

We now change to so-called isotropic coordinates in which
the spatial part of the metric is conformally flat. Let
$r = R- r_0$, so that the horizon lies at $r=0$.  Introducing
the Cartesian coordinates $x^i$ as usual on ${\bf R}^3$, we have

\begin{equation}
\label{isoRN}
ds^2 = - f^{-2} dt^2 +
f^2 \sum_{i=1}^3 dx^i dx^i,
\end{equation}
where $f = 1 + r_0/r$.  Similarly, the electro-magnetic potential
is given by $A_t = - f^{-1}$ with the spatial components
of $A$ vanishing.  As the function $f$ satisfies
Poisson's equation with a delta function source,

\begin{equation}
\label{1stP}
\partial^2_x f := \sum_{i=1}^3 \partial_i 
\partial_i f = - 4\pi  \delta^{(3)}(x),
\end{equation} 
the solution for the extreme black hole takes a form
similar to that seen in electrodynamics (except that the
Poisson equation is for the {\it inverse} of the electrostatic
potential).  Note that the relevant differential
operator is the Laplacian on a {\it flat} three-space and not
the one directly defined by the metric.  Such differential
operators will often appear below, and we will use the
convention that $\partial_x^2$ will always denote the flat-space
Laplacian associated with the coordinates $x$. Similarly, 
we will write $dx^2 := \sum_{i} dx^i dx^i$.

The analogy with electrostatics is quite strong. 
The above metric (\ref{isoRN}) and the associated electric field
in fact define the class of Majumdar-Papapetrou solutions \cite{MP}.
These are, in general, solutions of the Einstein-Maxwell
system coupled to extremal dust.  Recall that extremal
dust has the property that when two grains of dust are at rest,
their electrostatic repulsion is exactly sufficient to balance their
gravitational attraction and they remain at rest.
Modulo the conditions below,
any choice of the function $f$ in (\ref{isoRN}) yields a static solution
of the field equations corresponding to some distribution of
this dust.  For an asymptotically flat solution, we should
take $f$ to be of the form $1 + Q/r$ near infinity.  The one
restriction on $f$ is that
$\rho = -\frac{1}{4\pi} \nabla^2 f$ must be everywhere positive.
In particular, we 
will take it to be of the form
$\rho_0 + \sum_{k =1}^N r_k \delta(x-x_k)$ where $\rho_0$
is continuous.  The density (defined with respect to the
Cartesian coordinate system $x_i$) of extremal dust is given
by $\rho_0$ and each delta function will result in the presence
of an extremal black hole with charge radius $r_k$.  
In particular, near $x = x_k$, the
metric takes the same form as for an isolated extremal black hole.

Extremality is quite important for the simple form of this class of solutions.
It is only in the extremal limit that the repulsion induced by the
electric charge can `cancel' the gravitational attraction so that the
solution can remain static. If one adds any additional energy to the
solution, the non-linearities of gravity become more directly manifest.

Note that the source in (\ref{1stP}) lies
at the origin of the $x$-coordinates; i.e., at the horizon
of the black hole.  However, since the horizon of the black hole
is in fact not just a single point in space, $x=0$ is clearly
a coordinate singularity.  This means that although the support of the
delta function lies at $x=0$, this should not be interpreted
as the location of the black hole charge.  Rather, the role of
this delta function is to enforce a boundary condition on the
electric flux emerging from the black hole so that the hole
does indeed carry the proper charge. 

Of course, in 3+1 dimensions, we can also have magnetically
charged black holes.  In fact, we can have dyons, carrying
both electric and magnetic charge.  The corresponding
extremal solutions are given directly by electro-magnetic
duality rotations
of the above solution.

For future reference we note that there is a similar
set of solutions in 4+1 dimensions, though black holes
in five dimensions can carry only electric charge.
They take the form
\begin{equation}
\label{5RN}
ds^2 = - f^{-2} dt^2 +
f \sum_{i=1}^4 dx^i dx^i = -f^{-2} dt^2 + f dx^2,
\end{equation}
where $\partial_x^2 f = -2 \omega_3 (\rho_0 + \sum_{k=1}^N r_k^2
\delta(x-x_k))$, $\omega_3$ is the volume of the unit
three-sphere, $\rho_0$ is the
charge per unit $d^4x$ cell, and $r_k$ is the charge radius of the $k$-th
extremal black hole.  The fact that $r_k^2$, as opposed to $r_k$,
appears as the source
reflects the fact that the
fundamental solution of Poisson's equation in four dimensions is of the form
$r^{-2}$.

As we have
already commented, there is a coordinate singularity at
the black hole horizon.  Thus, the isotropic form of the metric
does not allow us to see to what extent the black hole, or even
the horizon, is non-singular.  However, if the black hole
is to have a smooth horizon, then a necessary condition
is that the horizon have non-zero (and finite) area.
That this is true of the above metrics is easy to read
off from (\ref{isoRN}) and (\ref{5RN}) by realizing that the 
divergence of $f^2$ or $f$ cancels the $r^2$ factor
that arises in writing $dx^2 = dr^2 + r^2 d\Omega^2$ in 
spherical coordinates.  While this is certainly not
a sufficient condition for smoothness of the horizon, it
will serve as a useful guide below.

Finally, for completeness, we display the conformal
diagrams for these solutions.  They are, in fact, identical
except for the dimension of the (suppressed) spheres of symmetry.

\vbox{
\centerline{\epsfbox{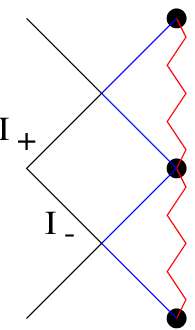}}
\centerline{Fig. 2. Conformal diagram for extreme Einstein-Maxwell 
Black Holes.}}

\medskip

Here $I_+,I_-$ denote the past and future null infinities of a particular
asymptotic region and the wavy line on the right
denotes the (timelike) singularity.
The black circles mark the ``internal infinities.''  These points
lie at an infinite affine parameter along any geodesic (spacelike, 
timelike, or null) from the interior.  This is often
referred to as the ``deep throat'' of the black hole.
The reason for this should be clear from the 
sketch below showing the embedding of a spacelike slice at
constant Killing time into flat space. 

\vbox{
\centerline{\epsfbox{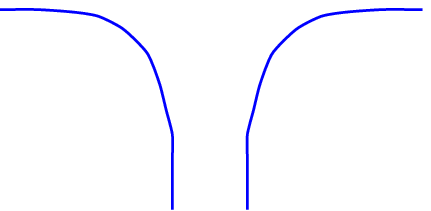}}
\centerline{Fig. 3. Embedding diagram for a Killing slice of the extreme
Einstein-Maxwell Hole.}}

\medskip

\subsection{Brane solutions in eleven dimensions}
\label{branes}

There are four (basic) solutions of eleven dimensional
supergravity that are of particular importance in string/M-theory.
These are known as the M-theory wave (basically the eleven-dimensional
version of the Aichelburg-Sexl metric \cite{AS}), the
M2-brane \cite{M2sol}
(electrically charged under $A_3$), the M5-brane (magnetically
charged under $A_3$), and the eleven dimensional
version of the Kaluza-Klein monopole \cite{Rafael,GP}.

Below, we discuss only the extremal versions of the solutions.
The non-extremal forms of the M-branes may be found in, e.g., \cite{GHT}.
In familiar 3+1 Einstein-Maxwell gravity, we are used to thinking
of extreme black holes as being some sort of marginal and perhaps
unphysical case.  Indeed, it is an important part of black hole
thermodynamics that one cannot by any finite (classical) 
process transform a non-extreme black hole into an extreme black hole.
Moreover, a real astrophysical black hole will quickly loose its
charge due to interactions with the interstellar medium.  Even in a pure
vacuum, quantum field theory effects in the real world cause black holes
loose their charge and to evolve toward neutral black holes.  
to loose mass.  However, this last statement is a consequence of the large
charge to mass ratio of the electron.  In a supersymmetric theory
of the type we discuss here, the charge of any particle is bounded
by its mass through a BPS bound so that objects like the familiar
electron do not exist\footnote{One may wish to ask what
relevance such supersymmetric theories can have to the real world.
The hope is that supersymmetry represents a `broken symmetry' of physics, so
that studies of supersymmetric theories can be relevant at some fundamental
level even though the connection with everyday physics is rather
complicated.}.  As a result, in a supergravity theory BPS black holes
in a vacuum do not discharge.  In fact, 
non-extreme black holes decay through the Hawking process {\it toward}
extremality and, due to quantum effects, one expects that
any non-extreme black hole will decay
to an extreme black hole in a very large but finite time.  
Thus, extreme black holes are of central importance in understanding
supersymmetric theories as they represent stable `ground states'
for black holes.

The extremal forms of the basic solutions are BPS, and in particular they
each have 16 Killing spinors, preserving half of the supersymmetry.
It is often said that an arbitrary BPS solution can
be built from these basic solutions.  To understand the
sense in which this is true, recall that a BPS solution
is extremal, and so carries charge.  BPS solutions are
classified in terms of the charges they carry, and 
the above `basic' solutions are in one-to-one correspondence
with the types of charge present in eleven dimensional
supergravity.  Since the charges are additive, one is tempted
to say that any solution with arbitrary amounts of the various charges
can be built up by
`combining' these basic solutions.  We will even see
that certain simple solutions carrying multiple charge
are in fact built from the basic solutions in
a simple way.  However, there is as yet no known method
for writing down a general BPS solution at all, much less
in terms of the basic solutions.  

Nevertheless, it is 
these four basic solutions (and those which are
built from them simply) which we will study here, leaving
the more complicated cases for the literature
(see in particular \cite{FS,pertdiv} for what is known
about the supergravity solutions corresponding to more complicated cases). 
Although it may not be obvious from the names, all four of the basic
solutions are associated with branes in string/M-theory.

Let us begin with the most obviously brane-like of the cases, 
the BPS M2- and M5-branes.   These are straightforward supergravity
analogues of the extreme Reissner-N\"ordstrom black holes of
Einstein-Maxwell theory\footnote{Interestingly, the
global structure of the {\it non-extreme}
M2- and M5-brane solutions is much like that  
of the Schwarzschild black hole, as opposed to that
of non-extreme Reissner-N\"ordstrom.  In particular, there is no
inner horizon and the singularity is spacelike as opposed to timelike.}.  
There is a corresponding notion of
isotropic coordinates in which the multi black hole 
solutions are given by solving a flat space Poisson equation
with delta-function sources.  The solutions of this Poisson equation
are typically
denoted $H_2$ for the M2-brane and $H_5$ for the M5-brane
and are referred to as `Harmonic' functions.  The details are different
for the two branes, but both should seem quite familiar from our review
of the Majumdar-Papapetrou solutions.

For the M2-brane, we introduce a set of three coordinates
$x_{\parallel}$ which should be thought of as labeling
the directions along the brane, and a set of eight coordinates
$x_\perp$ which should be thought of as labeling the
directions orthogonal to the brane.  As one of the 
$x_{\parallel}$ directions is the time direction, 
we define $dx^2_{\parallel} = - (dx_\parallel^0)^2 +
(dx_\parallel^1)^2 + (dx_\parallel^2)^2$.
The solution takes the form:

\begin{eqnarray}
\label{M2}
A_3 &=& - H_2^{-1} dt \wedge dx_{\parallel,1} \wedge dx_{\parallel,2} 
\cr
ds^2 &=& - H_2^{-2/3} dx_{\parallel}^2 + H_2^{1/3} dx_{\perp}^2
\end{eqnarray}
with $\partial_\perp^2 H_2$ equal to a sum of delta-functions.
Note that, near the delta function source, $H_2$ will diverge
like $r^{-6}$, where $r$ is the $x_\perp$ coordinate distance from the
source.  As a result, $H_2^{1/3}$ diverges like
$r^{-2}$, and the sphere at the horizon will
have non-zero (finite) area.   This suggests that the
horizon of the BPS M2-brane is smooth, and a careful
investigation \cite{GHT} does indeed show that this is
the case.  This is rather interesting, as the extremal
limits of black branes in lower dimensional supergravity theories tend, because
of the dilaton, to have singular horizons.  The 
global structure of the M2-brane is in fact much like
that of the extreme Reissner-N\"ordstrom black holes discussed
above.  The conformal diagram is just that of Fig. 1,
except that each point on the diagram now represents
a surface with both the topology and metric of 
${\bf R}^2 \times S^7$ instead of just a sphere. 

For the M5-brane, we introduce a set of six coordinates
$x_{\parallel}$ along the brane, and a set of five coordinates
$x_\perp$ orthogonal to the brane.  Again, the $x_\parallel$
directions include the time $t$.
The solution takes the form:

\begin{eqnarray}
\label{M5}
dA = F &=& - \frac{1}{4!} \partial_{x_\perp^i} H_5 \epsilon^{ijklm}
dx^j \wedge dx^k \wedge dx^l \wedge dx^m
\cr
ds^2 &=& - H_5^{-1/3} dx_{\parallel}^2 + H_5^{2/3} dx_{\perp}^2,
\end{eqnarray}
with $\partial_\perp^2 H_5$ equal to a sum of delta-functions. 
The different form of the gauge field as compared with 
(\ref{M2}) is associated with the
fact that this solution carries a magnetic charge instead of 
an electric charge.  Now the field $H_5$ diverges
at a delta-function source as $r^{-3}$, so that $H_5^{2/3}$
diverges like $r^{-2}$ and again the area of the spheres
is finite at the horizon.  Once again, a detailed
study shows that the horizon is completely smooth.
In fact, it turns out \cite{GHT} that this solution is smooth
{\it everywhere}, even inside the horizon!  Its
conformal diagram is rather different from those we have
encountered so far and is shown below.  The regions
marked A and B below (`in front of' and `behind') the horizon
are exactly the same.  In familiar cases, the singularity
theorems guarantee that something of this kind does not 
occur: compact trapped surfaces imply a singularity
in their future \cite{HawkingEllis}.  However, 
the fact that we deal with a black brane, and not
a black hole, means that the trapped surfaces are not in
fact compact.  The point here is that the horizon
is extended in the $x_\parallel$ directions.  What happens
when the solution is toroidally compactified by making 
identifications in  the $x_\parallel$ coordinates is an
interesting story that will be discussed below.

\vbox{\centerline{
\epsfbox{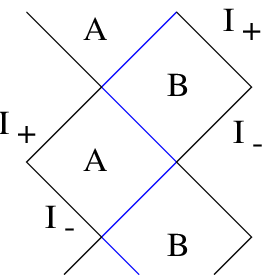}}
\centerline{Fig. 4.  Conformal diagram for the extreme M5-brane.}}

\medskip

The remaining two solutions have the interesting property
of being BPS despite the fact that the gauge field $A_3$
is identically zero.  This is not really a contradiction to
the condition of extremality when one notes (see section \ref{SUGRAVKK})
that under Kaluza-Klein reduction a momentum can act like
a charge.   Another useful perspective results from recalling that
two parallel beams of light (or two parallel gravitational waves) do not
interact gravitationally.  The same is true for any null particles.
Thus, one may say that the spatial components
of the momentum provide a gravitational repulsion and that the case
of null momentum is like the case of extremal charge, where this
repulsion just exactly balances the gravitational attraction due to the
energy of the particles.

The M-theory wave carries just such a
null momentum.
This solution was originally constructed \cite{AS} (in 3+1
dimensions) by boosting a Schwarzschild solution while
rescaling its mass parameter $M$ in order to keep the total
energy $E$ finite in some asymptotic frame.  This explains
the null momentum of the resulting solution.  It too can
be described in terms of a `harmonic function' $H_W$.
The M-theory wave may be thought of as the gravitational
field of a null {\it particle}, such as a graviton or a quantum of the
$A_3$ field in the
short wavelength (WKB) approximation.  We introduce a time
coordinate $t$, a coordinate $z$ in the direction of motion
of the particle, and a set of nine additional coordinates $x_\perp$.
In isotropic coordinates, the solution takes the form
\begin{eqnarray}
\label{Mwave}
ds^2 = - dt^2 + dx_\perp^2 +dz^2 +  (H_W-1) (dt-dz)^2
\end{eqnarray}
where $H_W(x_\perp)$ is a solution of $\partial_\perp^2 H_W =
-7 \omega_9 \rho$, 
where $\rho$ is again a source and $\omega_9$ is the volume of the unit
9-sphere.  When $\rho$ is
a delta-function, this solution is in fact singular at 
the source.  

Let us now turn to the Kaluza-Klein monopole.
This solution was originally constructed \cite{Rafael,GP} by using
the fact that the metric product of any two Ricci
flat spaces is Ricci flat.  Thus, one can make a static
solution of 4+1 Einstein gravity out of any solution to
four-dimensional, Euclidean gravity.  Such a solution is
Ricci flat, so the metric product with a line is also 
Ricci flat.  The metric product of Euclidean Taub-NUT
space \cite{NUT} with a line gives the 4+1 Kaluza-Klein monopole.
The eleven dimensional solution of interest here is simply
the metric product of Euclidean Taub-NUT space with
a 6+1 Minkowski space.  Recall that the Taub-NUT
solution is not asymptotically flat.  Rather, it is asymptotically
flat in three directions (to which we assign coordinates 
$x_{\perp}$) and the fourth direction (which we will call $\theta$)
is an angular coordinate for which the associated $S^1$
twists around the two-sphere to make a non-trivial asymptotic
structure.
Introducing coordinates $x_\parallel$
on the 6+1 Minkowski space, the solution takes the form:
\begin{equation}
\label{11KK}
ds^2 = dx_\parallel^2 + H_{KK} dx_\perp^2 + H^{-1}_{KK} (d\theta
+ A_i dx^i_\perp)^2
\end{equation}
with, of course, $A_3=0$.  Again, $H_{KK}$ satisfies
an equation of the form $\partial_\perp^2 H_{KK}= - 4 \pi \rho$ and
$a_k$ is determined from $H_{KK}$ via
$\partial_{x_\perp^i} H_{KK} = \epsilon_{ijk} \partial_{x_\perp^j} a_k$.
As usual, we find a coordinate singularity at the location of
the delta-function sources.

The story of this singularly
is just that of Taub-NUT space.  Suppose that $\theta$ is periodic with
period $L$.  Then the spacetime is in fact smooth
in the neighborhood of a `source' of the form 
$\rho= \frac{L}{4 \pi} \delta^{(3)}(x_\perp)$; in this case, 
(\ref{11KK}) actually represents a smooth geodesically complete
solution to the {\it source-free} 10+1 Einstein equations.  A
source of this sort
is referred to as a monopole of unit charge.  A multi-center 
solution\footnote{One with several delta-function sources.} 
is smooth whenever each separate center has this charge.
Now, if we take a limit of a multi-center solution in which
several of the centers coalesce into a single center with charge
greater than one, the resulting spacetime has a timelike
singularity at the source.  However, this singularity (with an integer
$n$ number of units of the above fundamental charge) has a particularly
simple form. It is a quotient of flat space, and in this
sense it is a higher dimensional dimensional version of a conical
singularity.

A favorite topic to include in discussions of black holes
is that of black hole entropy.  It is therefore natural to ask about
the entropy of the branes that we have discussed above.  Similar thermodynamic
arguments hold for black branes as for black holes, suggesting that
one should associate an entropy of $A/4G_{11}$ with such objects where
$A$ is the area (volume) of the horizon and 
$1/16\pi G_{11} = 1/2 \kappa_{11}^{-2}$ 
is the
coupling constant that stands in front of the supergravity action.
However, the Kaluza-Klein monopole and M-theory wave
have no Killing horizons and so presumably
carry no such entropy.  The M2- and M5-branes are a bit more subtle.
On the one hand, their horizons are homogeneous surfaces that are non-compact.
As such, one might be tempted to assign them infinite entropy.
Some further
insight into the issue is gained by using the fact that the solutions
are invariant under translations in the spatial $x_{\parallel}$ coordinates
to make toroidal identifications and compactify the horizons.
We can then calculate the horizon area and, because the norms
of the Killing fields $\partial_{x_\parallel}$ vanish on the horizon, 
the result is zero.  Thus, at least when compactified in this way, 
the M2- and M5-branes also carry no entropy.  This is another sense
in which such solutions are `basic.'

One might guess that there is something singular about the zero-area horizons
of the compactified M2- and M5-branes.  However, since those solutions
were constructed by making discrete identifications of spacetime with
smooth horizons, the curvature and field strength cannot diverge
at the zero-area horizon.  It turns out that 
the situation is essentially the same as that which
arises \cite{BTZ} when $AdS_3$ is identified to make the $M=0$ 
BTZ black hole.  The 
initially spacelike Killing
fields $\partial_{x_\parallel}$ become null on the horizon but also have
fixed points.  Thus, the horizon of the compactified 
solution has both closed null curves and a `Lorentzian conical singularity.'

\subsection{Brane Engineering}
\label{smear}

Before leaving eleven dimensions, a few words are in order
on two of the basic techniques in `Brane Engineering,' constructing
new brane solutions from old.  The particular techniques
to be discussed are known as smearing and combining
charges.

Smearing is particularly straightforward.  It is based
on the observation that each type of `basic' solution
above is related to the solution of a linear differential
equation.  Using a delta-function source gives a solution
which preserves some set of translation symmetries (in the 
parallel directions) and breaks another set (in the $x_\perp$
directions).  However, a solution can be obtained that
preserves more translational symmetries by using 
a more symmetric source, e.g., one supported on a line, 
plane, or a higher dimensional surface.  Constructing such
a solution can be thought of as `smearing out' the charge
of a less symmetric solution.   Smearing out
a given brane solution often results in a spacetime with 
a singular horizon.  However, this need not be especially
worrying if one regards the smeared solution as merely
an effective description analogous to describing a collection of
discrete atoms as a continuous fluid.  One imagines an array of branes in which
a large number of unsmeared basic
branes are placed in the spacetime with a small spacing between
the branes.   We will soon
see that smearing is an important step in the construction of
BPS black brane solutions with finite entropy.

The next technique to discuss is that of combining the basic
types of charge.  As mentioned above, this is in general
rather difficult.  If, however, two solutions preserve some of the same
supersymmetries {\it and} they have been
engineered to have the same translation symmetries (for example, by
smearing), then they tend to be rather easy to combine.
Making a simple guess as to the way in which the
relevant harmonic functions ($H_2,H_5,H_W,H_M$) should enter the metric and
gauge fields tends to lead to a solution to the supergravity equations
which preserves the common supersymmetries. 

So far as I know, there are no general theorems available on this subject.
We will thus content ourselves with a few simple examples.
We have already discussed the solution (\ref{M2}) corresponding to
a set of parallel M2-branes.  This solution preserves half of the
original 32 supersymmetries of 10+1 supergravity.  The particular
supersymmetries that are broken are related
to the plane in space along which
the M2-branes are oriented.   Let us call the spatial coordinates along
these branes $x^1_\parallel$ and $x^2_{\parallel}$.
We could also consider another set
of M2-branes oriented along another plane associated with two other
coordinates $y^1_{\parallel}$ and $y^2_{\parallel}$, which are to
be orthogonal to the $x_\parallel$ coordinates.  A set of  solution containing
both types of branes and preserving the $8$ supersymmetries common to
both sets of M2-branes separately is given by:

\begin{eqnarray}
\label{2M2}
A &=& - H_x^{-1} dt \wedge dx_{\parallel,1} \wedge dx_{\parallel,2}  
- H_y^{-1} dt \wedge dy_{\parallel,1} \wedge dy_{\parallel,2}   
\cr
ds^2 &=& - H_x^{-2/3} H_y^{-2/3} dt^2 +  H_x^{-2/3} H^{1/3}_y
 dx_{\parallel}^2 + H_x^{1/3}  H_y^{-2/3} dy_{\parallel}^2
 \cr &+& H_x^{1/3} H_y^{1/3} dx_{\perp}^2
\end{eqnarray}
where $H_x$, $H_y$ are functions only of the six spatial coordinates $x_\perp$
that are transverse to both sets of branes.  The functions $H_1$ and
$H_y$ are, as usual, related to source distributions through
$\partial^2_{x_\perp} H_x =-7 \omega_8 \rho_x$ and 
$\partial^2_{x_\perp} H_y = - 7 \omega_8 \rho_y$ and the distributions $\rho_x$
and $\rho_y$ may be arbitrary functions of $x_\perp$.    

Note
that the form (\ref{2M2}) is just like that of (\ref{M2}) except that
we include two Harmonic functions.  A given term in the metric (\ref{2M2})
is multiplied by a power of each harmonic function determined 
by whether the term refers to distances along or transverse to the
corresponding brane.  These powers are identical to the ones in (\ref{M2}).

In the solution (\ref{2M2}), we
have taken the two sets of branes to be completely orthogonal to each other.
However, other choices of the relative angle still preserve the same
amount of supersymmetry.  If one thinks about the coordinates
$x_\parallel$, $y_\parallel$ as two holomorphic coordinates on ${\bf C}^2$, 
then the requirement for a supersymmetric solution is that 
the $x_\parallel$ and $y_\parallel$ planes are related by a $U(2)$
transformation \cite{angles}
as opposed to a more general $O(4)$ transformation.  
The metric in this case takes a similar form, with the part of the
metric on the four-space spanned by $x_\parallel$, $y_\parallel$
taking a certain Hermitian form.

Combining the two sets of branes without first smearing them to
generate four translation symmetries is, however, rather more difficult.
It turns out that, when one or both of the sets of branes is `localized' (i.e., 
not completely spread out along the other set of branes) then 
the supergravity equations no longer cleanly divide into pieces
describing each set of branes separately.  The case where only
one set is localized (and two translational symmetries remain)
is still tractable, however.  The solution
still takes the same basic form (\ref{2M2}) and 
construction of the solution
still splits into two parts.  
One can first solve a standard flat space
Poisson equation for the Harmonic function
$H_x$ associated with the delocalized set of branes. One then
has a linear differential equation to solve for the localized brane 
Harmonic function $H_y$, where
$H_x$ appears in the particular differential operator to be inverted.
When the
separation (in the $x_\perp$ directions) of the two sets of branes
vanishes, this `interaction' between the two sets of branes generates some
interesting effects related to black hole no-hair theorems 
\cite{SS,AP,pertdiv}, see \cite{mont} for a discussion
geared to an audience of relativists.  Localizing both branes
requires the solution of a non-linear partial differential
equation (see \cite{FS} for discussion of
the related M5-brane case).  Although their solution is not yet understood, it
appears \cite{pertdiv} that it will be rather far from the simple
structure associated with the basic branes.  

Let us now return to the smeared solution (\ref{2M2}) and consider
the case in which
$\rho_x = \rho_y = r_0^4 \delta^{(6)}(x_\perp)$.  We then find that $H_x$ and
$H_y$ diverge at $x_\perp = 0$ like $|x_\perp|^{-4}$.  As a result, 
the 5-spheres at $x_\perp=0$ are infinite in volume.  Thus, this
solution is somewhat singular.  However, adding a third M2-brane in
another completely orthogonal ($z_\parallel^1, z_\parallel^2$) plane
yields a non-singular solution. 
The metric and gauge field

\begin{eqnarray}
\label{3M2}
A &=& - H_x^{-1} dt \wedge dx_{\parallel,1} \wedge dx_{\parallel,2}  
- H_y^{-1} dt \wedge dy_{\parallel,1} \wedge dy_{\parallel,2}   
- H_z^{-1} dt \wedge dz_{\parallel,1} \wedge dz_{\parallel,2}   
\cr
ds^2 &=& - H_x^{-2/3} H_y^{-2/3} H_z^{-2/3}
 dt^2 +  H_x^{-2/3} H^{1/3}_y H_z^{1/3}
 dx_{\parallel}^2 + H_x^{1/3} H_y^{-2/3} H_z^{1/3} dy_{\parallel}^2 \cr
&+& H_x^{1/3} H_y^{1/3} H_z^{-2/3} dz_{\parallel}^2 
 + H_x^{1/3} H_y^{1/3} H_z^{1/3} dx_{\perp}^2
\end{eqnarray}
for $\partial^2_{x_\perp} H_{x,y,z} (x_\perp) = -2 \omega_3
\rho_{x,y,z} (x_\perp)$ 
yields a BPS solution
of the supergravity equations that preserves 1/8 of the supersymmetry
(i.e., 4 supercharges) and has a smooth horizon.  Moreover, this solution
has the property that the translational Killing fields 
$\partial x^i_{\parallel}$,
$\partial y^i_{\parallel}$,
$\partial z^i_{\parallel}$ have norms that do not vanish on the horizon.

In contrast, recall that while the solution (\ref{M2}) for a single
M2-brane has a smooth horizon, the spatial translational Killing fields
have vanishing norm there.  As mentioned above, this means that compactifying
a single M2-brane by, for example, taking the coordinate $x_{\parallel}^1$
to live on a circle, yields a solution with a conic singularity at
the horizon and vanishing entropy.
On the other hand, because the norms of the spatial translations
do not vanish for the solution (\ref{3M2}), it compactifies nicely
into a black object with finite horizon area.  
This is the simplest BPS black brane solution with a finite entropy
and, as a result, it is the simplest solution for which
a microscopic accounting of the entropy has been given in string theory.
A straightforward calculation shows the the horizon area is
\begin{equation}
\label{3M2area}
A = \omega_3 r_x r_y r_z L_{1x}L_{2x}L_{1y}L_{2y}L_{1z}L_{2z},
\end{equation}
where $\omega_3$ is the volume of the unit three-sphere and
the $L$'s are the lengths of the various circles on which the solution
has been compactified.

Now, charges are quantized in string/M-
theory and it is useful to express the entropy in terms of the number
of charge quanta $Q_x,Q_y,Q_z$ carried by the various branes.  
The tension of a single M2-brane is $(2\pi)^3l_p^{-3}$, where $l_p$ is the
eleven-dimensional Plank length, defined by
$16 \pi G_{11} = 2 \kappa_{11}^2 = (2\pi)^8 l_p^9$.  
Note that $r^2_x$ is a measure
of the charge {\it density} of the $x$-type branes per unit cell of the $y,z$
four-space. As such, $r^2_x$ is proportional to $Q_x/L_{1y}L_{2y}L_{1z}L_{2z}$.
Inspection of the area formula (\ref{3M2area}) thus shows that
rewriting the area in terms of the integer charges will remove
the factors of $L$.  Putting in the proper normalization coefficients, 
the result turns out to be
\begin{equation}
\label{BHE}
A/4G_{11} = 2\pi \sqrt{ Q_x Q_y Q_z}.
\end{equation}
We will comment briefly on the corresponding microscopic counting of
states in section \ref{SandD}.

\section{Kaluza-Klein and Dimensional Reduction}
\label{KKred}

So far, we have dealt almost exclusively with branes in
eleven dimensional supergravity.  We focused on this case
for two reasons.  The first reason is that eleven is the maximal number
of dimensions in which there is a supergravity theory
with fields of spin less than or equal to two. As such,
it seems to play a role of fundamental importance in string/M-theory.
Most other lower dimensional supergravity theories can
be obtained through the Kaluza-Klein mechanism 
in which some subset of the dimensions are taken to be compact
and small.  This mechanism is discussed in sections  
\ref{KK} and \ref{SUGRAVKK} below.

The second reason is that 
supergravity is in fact simpler in eleven dimensions than in
lower dimensions.  The eleven dimensional theory contains
only two bosonic fields, the metric and the 3-form, and one
fermionic field.  In particular, there is no dilaton and the
3-form is minimally coupled.  In contrast, supersymmetry in 
lower dimensions forces the theory to contain a plethora of
different bosonic fields, including a dilaton.  It is the
dilaton and its cousins, the other ``moduli,'' which make 
classical supergravity in ten dimensions or less so much
different from familiar Einstein-Maxwell theory.  This is because
the dilaton couples non-minimally to the various
fields, with the result that different fields couple to what are
effectively different metrics.  In other
words, the equivalence principle is violated.  The various
metrics share the same conformal
structure and are related by different factors of $e^\phi$, where
$\phi$ is the dilaton.  Thus, by focusing first on the eleven
dimensional case, we have been able to make maximal use of
our 3+1 Einstein-Maxwell intuition.

However, it turns out that supergravity in 9+1 dimensions (just one
dimension down from the maximum) has an important property
that does not follow just by thinking of it as the compactification
of a 10+1 theory:  The 9+1 theories admit a self-consistent
perturbative quantization in terms of strings\footnote{It should be
mentioned that string theory is not a quantization of
pure 9+1 supergravity; string theory modifies the physics
even at the classical level.  See section \ref{bfe} for (a few) more
details.}.  This means 
that the powerful technology of perturbative quantum field
theory can be brought to bear on questions concerning their
quantum dynamics.  This perturbative technology can in particular
be applied to certain branes in 9+1 supergravity.
It is through this fusion of supergravity and
perturbative field theory that
string/M-theory has been revolutionized in recent years
via studies of duality, black hole entropy, and more recently
the Maldacena conjecture or AdS/CFT correspondence.

This article is not the place to enter into
a detailed discussion of string perturbation theory, hough we will
comment briefly on the subject in section \ref{pert}.
The reader interested in learning that subject should
consult the standard references \cite{Joe,GSW}.
Our purpose here is to provide a clear
picture of the supergravity side of BPS brane physics, and in particular
to discuss their relationship with the eleven dimensional
theories.  Thus, we begin with a discussion of Kaluza-Klein
compactification in non-gravitational theories.  We then discuss 
in detail the Kaluza-Klein reduction of eleven-dimensional supergravity
in the presence of a small $S^1$, which yields so-called type IIA
supergravity in 9+1 dimensions.  This sets the stage for our discussion
of 9+1 branes in section \ref{10branes}.

\subsection{Some remarks on Kaluza-Klein reduction}

\label{KK}

The idea of the Kaluza-Klein mechanism is that, at low
energies,
a {\it quantum} field theory on an $n+d$-dimensional
spacetime in which $d$ of the dimensions are compact
behaves essentially like a quantum field theory on
an $n$-dimensional spacetime.  To see why, consider
a free scalar field on $M^n \times S^1$ where $M^n$ is
$n$-dimensional Minkowski space.  Let us consider the mode
spectrum of the scalar field.  Modes are labeled by
an $n$-vector momentum $p$ and an integer $k$ corresponding
to momentum around the $S^1$.  Suppose that
the length of the $S^1$ is $L$, 
so that the dispersion relation associated with one-particle
excitations is $E^2 = p^2 + (k/L)^2$.  If we now consider
the theory at energy scales less than $1/L$, the only
states with such a low energy are states with $k=0$; i.e., 
states that are translationally invariant around the $S^1$.

In this way, our scalar field reduces at such energy scales
to a quantum field on $n$-dimensional Minkowski space.
Note that this is an intrinsically quantum mechanical effect.  Note also that
it is associated with the discrete spectrum of the Laplacian on
a circle.  Since the Laplacian has a discrete spectrum on any
compact space, the same mechanism operates with any choice
of compact manifold.    The simplest cases to analyze
are those in which the spacetime is a direct product of
a non-compact spacetime $M$ with a compact manifold $K$, and
in which $K$ is a homogeneous space.
In that case, the lower dimensional (reduced) theory is obtained
from the higher dimensional one simply by taking the fields
to be invariant under the symmetry group that acts transitively on the compact
manifold.   On a general manifold of the form $M \times K$, 
the reduced theory is given by considering the zero-modes
of the Laplacian (or other appropriate differential operator) on
$K$.  Similar, but less clean, mechanisms
may apply even when the spacetime is not a direct product of a compact
and a non-compact spacetime. 

The effect of compactification on interacting fields is similar.
At the perturbative level the story is exactly the same, and
non-perturbative effects seldom change the picture significantly.

\subsection{Kaluza-Klein in (super)gravity}
\label{SUGRAVKK}

Let us now consider this mechanism in a theory with gravity.
Since the spacetime metric is dynamical, this case is perhaps
not as clean cut as the scalar field example just discussed.
However, at the perturbative level, one may treat
gravity just as any other field.  To this extent then, the
same conclusions apply.  Also, our general experience with
quantum mechanics and the uncertainty principle
makes it reasonable on more general grounds
to expect that excitations
associated with the small compact space will be expensive in
terms of energy.  Thus, at least at first glance, we expect
that gravity on a manifold of the form $M \times K$ 
reduces at low energies to a theory on the non-compact
manifold $M$.  

The case in which we are most interested is the Kaluza-Klein
reduction of eleven-dimensional supergravity to 9+1
dimensions. 
We expect the reduced theory to be obtained by considering
the class of eleven-dimensional field configurations that
are translationally invariant around the $S^1$.  Let us therefore
assume that our eleven dimensional spacetime ${\cal M}$
has a spacelike Killing
vector field $\lambda^\mu$ whose orbits have the topology $S^1$.
It is convenient to normalize
this Killing field to have norm $+1$ at infinity 
and to denote the length of the Killing orbits there by $L$.
The Killing field is not necessarily
hypersurface orthogonal.

Since the translation group generated by the
Killing field acts nicely (technically, `properly
discontinuously' \cite{HawkingEllis} on our eleven dimensional
spacetime ${\cal M}$, we may consider
the quotient of the smooth topological space
${\cal M}$ by the action of this group.
The result is a new topological space $M$, which is a ten
dimensional smooth manifold.  This is the manifold on which our
$10=9+1$ dimensional reduced theory will live.  

By using the metric, 
we define a set of projection operations on the various 10+1
fields with each projection providing a different field in the 9+1
dimensional spacetime.
Recall that a field is an object which transforms in a certain
way under local Lorentz transformations (i.e., diffeomorphisms) of the
manifold.  The diffeomorphisms of the 9+1 manifold will be that
subgroup of the eleven-dimensional diffeomorphisms that leaves the
killing field $\lambda^\mu$ invariant.  Thus, the transformations
that become the diffeomorphisms of the $9+1$ manifold are a proper 
subgroup of the 10+1 diffeomorphisms and a single 10+1 field
can contain several 9+1 fields.

To see how the 9+1 fields are constructed, consider any
coordinate patch $U$ (with coordinates $x^a$) on the 9+1 manifold $M$.  
If $V \subset {\cal M}$ is the preimage of $U$ under 
the above quotient construction, then each $x^a$ defines
a function on $V$.  Since no linear combination of the
gradients of the $x^a$ functions can be proportional to the Killing field
$\lambda^\mu$, we can complete this set of functions to
a coordinate patch\footnote{Technically, $\theta$ is not
quite a valid coordinate because it is periodic instead of single
valued.  The reader can easily fill in the appropriate details if
desired.} on $V$ by adding a coordinate $\theta$
which is proportional to the Killing parameter along any orbit of
$\lambda^\mu$; i.e., satisfying $\theta_{,\mu} \lambda^\mu 
= \lambda^\mu \lambda_\mu$. 

This coordinate system gives an explicit realization of the
natural decomposition of the 10+1 fields into a set of 9+1 fields.
The set of gradients $x^a_{,\mu}$ of the 9+1 coordinates
define a projection operation on any contravariant (upper) index, as does
the gradient $\theta_{,\mu}$ of the coordinate $\theta$.  Thus, 
from the 10+1 contravariant metric $g^{\mu \nu}$, we can define the
9+1 metric $g^{ab} =  x^a_{,\mu} x^b_{,\nu} g^{\mu \nu}$, 
a 9+1 abelian vector field $A_1^a = - x^a_{,\mu} \theta_{,\nu} g^{\mu \nu}$, 
and a 9+1 scalar field $\phi$ through 
$L e^{4\phi/3} = \lambda^\mu g_{\mu \nu} \lambda^\nu$.  The particular
coefficient of $\phi$ is chosen so that it is canonically 
normalized\footnote{When $2\kappa_{10}^2$ is set to one, see below.}.
This $\phi$ is the famous dilaton of string theory, and it is this field
which is responsible for many of the differences between
supergravity in less than eleven dimensions and familiar 
Einstein-Maxwell theory.
It is clear that all of these fields transform in an appropriate way under
9+1 diffeomorphisms.

There
are several important observations to make about these
definitions.  The first is that nondegeneracy of the 10+1
metric implies non-degeneracy of the 9+1 metric.  Thus, 
$g^{ab}$ has an inverse which gives the covariant metric $g_{ab}$.

The second is that the scalar has been defined by the
norm of the Killing field and not the norm of $\theta_{,\mu}$
as one might expect.  The point is that these two objects are related.
To see this, let us first note that the coordinates $x^a$
are constant along the orbits of the Killing field.
Thus, the Lie derivative of $x^a$ along $\lambda^\mu$ vanishes, and
we have $x^a_{,\nu} g^{\nu \mu} \lambda_{\mu} =0$.
This means that the gradients $x^a_{,\mu}$ span the space
orthogonal to $\lambda_\mu$ at each point.  But, by
definition, $\lambda^\mu \theta_{,\mu} = \lambda^\mu \lambda_\mu$.
Thus, we find that $\theta_{,\mu} - \lambda_{\mu}$ is of the
form $c_a x^a_{,\mu}$ where $c_a$ is some function on the 9+1
spacetime.  This 
fact, together with
the definition of $A_1$, can be used to derive the relation:
\begin{equation}
c_a = - g_{ab} A_1^b.
\end{equation}
Thus, we have 
\begin{equation}
\theta_{,\mu} \theta_,{}^\mu = \lambda^\mu \lambda_{\mu} + A_{1a}A_1^a.
\end{equation}  
We see that the definition of
$\phi$ differs from the seemingly more natural one only
by a function of the vector field $A_1$.  Choosing to
write $\phi$ directly in terms of the Killing field $\lambda^\mu$
removes a mixing between the vector field and scalar that
would otherwise obscure the physics.
Note that we have related the scalar field $\phi$ to the logarithm
of the norm of the Killing field, and that this norm is
positive by assumption.

Finally, let us consider the vector field $A_1.$
Although
we have $x^a_{,\nu} g^{\nu \mu} \lambda_{\mu} =0$,
the vector field $A_1$ need not vanish.  It represents
the twist of the gauge field; that is, the failure of the
gauge field to be hypersurface orthogonal.
Note that there is a freedom
to redefine the zero of $\theta$ at each value of the $x^a$.  This amounts
to the transformation $\theta \rightarrow \theta - \Lambda(x)$.  Under
this operation, we see that the $9+1$ metric $g^{ab}$ is not affected, 
and neither is the scalar (since it depends only on the norm
of the Killing field)
while the vector field transforms as $A_1^a \rightarrow A_1^a + 
\Lambda_{,b} g^{ab}$; i.e., $A_{1a} \rightarrow A_{1a} + 
\Lambda_{,a}$.  Thus, we see that $A_1$ is in fact an abelian
gauge field.  

It is interesting to ask about the charge to which this gauge field
couples, as the field itself arose directly from the reduction of the
gravitational field in eleven dimensions.  Let us therefore consider
the transformation generated by the electric charge, a global $U(1)$
gauge rotation.  As we have just seen, in terms of the higher dimensional
spacetime this is a translation along the Killing field $\lambda^\mu$.
It is therefore generated by a momentum.  Thus we see that the
charge to which the gauge field $A_1$ couples is nothing other than
momentum around the internal $S^1$ when viewed from the higher
dimensional perspective.  A timelike energy-momentum vector in the
eleven dimensional spacetime translates in the ten-dimensional
context into a charge and a ten-dimensional energy-momentum tensor
satisfying a BPS bound.

In performing calculations, it is often useful to express
the above decomposition in terms of the eleven dimensional
{\it covariant} metric $ds_{11}^2$.  The reader may check that
we have
\begin{equation}
ds_{11}^2 = g_{ab} dx^a dx^b + e^{4 \phi/3}[d\theta + A_{1a} dx^a]^2
\end{equation}

One might think it is natural to decompose, the antisymmetric 3-form 
into a 9+1 3-form $\hat A_{3}^{abc} = A_{3}^{\mu \nu \rho}
x^a_{,\mu}x^b_{,\nu} x^c_{,\rho}$ and and a 2-form 
$A_2^{ab} = A_{3}^{\mu \nu \rho}
x^a_{,\mu}x^b_{,\nu} \lambda_{\rho}$ in order that both be invariant
under the gauge transformations $A_1 \rightarrow A_1 + d \Lambda_0$.  
However, it turns out that the 3-form $\hat {A}_3$ 
transforms nontrivially under the gauge transformations associated
with the 2-form potential $A_2$.  The various gauge transformations
cannot be completely disentangled and in fact the standard
choice is to use
\begin{equation}
A_3 = \frac{1}{3!}  \tilde{A}_{3 abc} dx^a\wedge dx^b \wedge dx^c
+ \frac{1}{2!} A_{2ab} dx^a \wedge dx^b \wedge d\theta.
\end{equation}
The decomposition of the fermionic
fields is similar, but we will not go into this in detail.

The gauge symmetry
of the eleven dimensional $A_3$ implies that there are
independent 9+1 gauge symmetries $\tilde A_3 \rightarrow \tilde 
A_3 + d \Lambda_2$ and $A_2 \rightarrow A_2 + d \Lambda_1$ where $\Lambda_n$
are arbitrary $n$-forms.    The two form is invariant under $A_1 \rightarrow
d \Lambda_0,$ but we have $\tilde A_3 \rightarrow \tilde 
A_3 + A_2 \wedge d\Lambda_0$.
From here on, we drop the tilde ( $\tilde{}$ ) on $\tilde A_3$.
This mixing of gauge transformations leads to interesting phenomena involving
non-conservation of charge in the reduced theory, but this will not arise
in the simple spacetimes discussed below.  As a result, we will not go into
the details here.

\subsection{On 9+1 Dynamics: Here comes the dilaton}

The dynamics for the 9+1 theory follows from that of eleven dimensions
by inserting the
relations between the 9+1 fields and the 10+1 fields into the
action.  The result is an action principle for the 
9+1 theory which takes the form

\begin{eqnarray}
\label{10daction}
S_{9+1,bosonic} &=& \frac{1}{2\kappa_{10}^2} \int d^{10} x
\Bigl[
\sqrt {-g} \left(e^{2\phi/3} R - \frac{1}{2} e^{2 \phi} F_2^2 
\right) \cr
&-&
\frac{1}{4 \kappa^2_{10}} \int d^{10}x \sqrt{-g} 
\left( e^{-2\phi/3}F^2_3 + e^{2\phi/3}
\tilde F_4^2 \right) \cr
&-& \frac{1}{4\kappa_{10}^2} \int A_3 \wedge F_3 \wedge F_4 \Bigr].
\end{eqnarray}
where all quantities refer to the 9+1 dimensional fields, 
$F_n = dA_{n-1}$, and $\tilde F_4 = d A_3 - A_1 \wedge F_3$.
As opposed to $F_4$ itself, the
new field strength $\tilde F_4$ is in fact invariant under
the gauge transformations of the $A_1$ potential.
We have also defined $\kappa_{10}^2 = \kappa_{11}^2/L$.

One important feature of (\ref{10daction}) is that
the field $\phi$ appears all over the place, with 
different factors of $e^\phi$ appearing in different terms.
The upshot of this is that the various gauge fields do not
couple minimally to the metric $g$, but instead the action
includes derivative couplings between $\phi$ and the
gauge fields.  Now, we do in fact have the freedom to
mix the metric with $\phi$ by rescaling the metric
by some power of $e^\phi$.  This can be used to
make any one of the gauge fields couple minimally to the
new metric, or to remove the factors of $e^\phi$ in front
of the scalar curvature term, and put the action in a form
more like that of familiar Einstein-Hilbert gravity.  However, 
because of the way that different factors of $e^\phi$ appear
in the different terms, this cannot be done for all fields
at once.  Thus, we may think of each different gauge field 
as coupling to a different metric.  

A short calculation shows that the gauge fields $F_2$
and $F_4$ couple minimally to $e^{2\phi/3}g$ while the gauge
field $F_3$ couples minimally to $e^{-\phi/3}g$.
In doing this calculation, it is important to realize that
terms like $F_2^2$ contain implicit factors of the metric $g$
which has been used to contract the indices (see appendix).  
On the other
hand, it is for the `Einstein metric' $e^{\phi/6}g$
that the gravitational part of the action takes the standard
Einstein-Hilbert form (the integral of the scalar curvature
density) without any extra factors of $e^\phi$.  

The choice
of a particular metric in the class $e^{\alpha \phi} g$
is known as the choice of conformal frame.  One can make a choice
of frame that simplifies a given calculation, if one desires.
It is interesting to note that, in the 
conformal frame which follows from the Kaluza-Klein
reduction, the field $\phi$ has no explicit
kinetic term so that its variation leads to a constraint.
It turns out that this is just a combination of the usual
constraints that one would expect in a gravitating theory, 
and that a term of the form $\partial_a \partial^a \phi$ does
appear in the equations of motion obtained by varying the metric 
in that frame.

The two most useful choices of conformal frame are the
Einstein frame discussed above, and the so-called string frame.
The action in the Einstein frame is a handy thing to have
on hand, so we will write it down here.   If we
now let $g_E$ denote the metric in the Einstein frame and let
$R_E$ be the associated curvature, the
action is
\begin{eqnarray}
\label{Eaction}
S_{bosonic} &=& \frac{1}{2\kappa_{10}^2} \int d^{10}x
\sqrt {-g_E} \left(R_E - \frac{1}{2} \partial_a\phi \partial^a \phi\right)
\cr &-& 
\frac{1}{4 \kappa_{10}^2} \int d^{10}x  
\sqrt{-g_E} \left( e^{3 \phi/2} |F_2|^2 + e^{-\phi} |F^2|_3 + e^{\phi/2}
|\tilde F_4|^2 \right) \cr
&-& \frac{1}{4 \kappa_{10}^2} \int  A_2 \wedge F_4 \wedge F_4 .
\end{eqnarray}
Note that, 
in Einstein frame, the gauge fields are all sources for the 
dilaton but the metric is not.  Also, since the kinetic term for the
dilaton now take the standard form, we can see that the dilaton would
be canonically normalized if we set $2 \kappa_{10}^2$ to one.
Finally, since it is in this frame that the gravitational dynamics take
the familiar Einstein-Hilbert form, this is the frame in which the 
standard ADM formulas for energy and momentum may be applied.

The string frame is defined by taking the metric to be
$e^{2\phi/3} g$, where $g$ is the original metric from dimensional
reduction.  That is, the string metric $g_S$ and the Einstein
metric $g_E$ are related by  $ds_{E} = e^{-\phi/2} ds_{S}$.
It should not be a surprise that the string metric is useful
as two of the gauge fields ($F^2$ and $F^4$) couple minimally
to this metric.  These two gauge fields are known as
{\it Ramond-Ramond} (R-R) gauge fields while $F_3$ is known as
the {\it Neveu-Schwarz Neveu-Schwarz} (NS-NS) gauge field.  
For an explanation of how this terminology arose in string
perturbation theory, see \cite{Joe}.  The potential
$A_{2}$ for this field is commonly written $B_{ab}$ and when
string theorists discuss ``the B-field,'' it is this potential to which
they are referring.    

However, what makes the string metric especially useful
is that it turns out to be the metric to which fundamental
strings (which we have not yet discussed) couple.  For this
reason, it is in the string frame that one can make the
most direct contact with string perturbation theory.
This, however, is a discussion for another place and time.
Here, we wish only to record the metric in the string frame for
the reader's future use.

\begin{eqnarray}
\label{Stringaction}
S_{bosonic} &=& \frac{1}{2\kappa_{10}^2} \int d^{10}x
\sqrt {-g_S} e^{-2\phi} 
\left(R_S + 4 \partial_a\phi \partial^a \phi - \frac{1}{2}|F_3|^2
\right) \cr &-& 
\frac{1}{4 \kappa_{10}^2} \int d^{10}x  
\sqrt{-g_S} \left( |F_2|^2 + |\tilde F_4^2| \right) \cr
&-& \frac{1}{4 \kappa_{10}^2} \int  A_2 \wedge F_4 \wedge F_4 .
\end{eqnarray}
After setting $c=\hbar =1$, the parameter $\kappa^2_{10}$ has units
of $(length)^8$.  It is useful to write $\kappa_{10}^2 = (2 \pi)^6
g_s^2 l_s^8$ where $l_s$ is the ``string length'' and $g_s$ is the
``string coupling.''    For more on the separate role of
$g_2$ and $l_2$, see section \ref{pert}.

In the above, we have discussed only the compactification of eleven dimensional
supergravity on a circle.  One can, of course, consider further
compactifications to smaller dimensional manifolds.  The story in that
case is much the same except that the number of (lower-dimensional)
fields generated increases rapidly.  In particular, further compactification
generates large numbers of massless scalars that couple non-minimally
to the various gauge fields.  These cousins of the dilaton are generally
referred to as {\it moduli}.  

All of these moduli have a tendency
to diverge at the horizon of an extreme black hole, making the solution
singular.  One may think of the issue as follows:  the moduli, like
the dilaton, couple to the gauge fields so that the squared
field strengths $F^2$ act
as sources.  This can be seen from the action (\ref{Eaction}) in the
Einstein frame.  Non-singular extremal black hole solutions typically have
an infinite throat, as in the four and five dimensional Einstein-Maxwell
examples discussed earlier.  This means that a smooth such solution
would have an infinite volume of space near the horizon in which
the gauge field strengths are approximately constant.  Thus, unless
these gauge fields are tuned to have $F^2 =0$ or the various
gauge fields are somehow played off against one another, there
is an infinite source for the moduli.  
As a result, it requires some care to construct an
extremal black hole solution with a smooth horizon and such
solutions necessarily carry more than one charge.   
For a brane solution, the norm of each Killing field acts like
a modulus whose sources must be properly tuned.

This is essentially
the issue encountered at the end of section \ref{smear} in which
it was found that three charges (in the case, three different types of
M2-branes) were required to obtain a brane solution in which the norms
of the Killing fields did not vanish on the horizon.  Recall
that a Killing field with positive norm is required to Kaluza-Klein
reduce the spacetime to a solution of lower-dimensional supergravity.
Because the three charge solution (\ref{3M2}) has six
Killing vector fields whose norms do not vanish on the horizon, it
may be reduced all the way down to a solution of 4+1 gravity.
In this context, it represents an extreme black {\it hole}.
In fact, it reduces to just the standard 5+1 extremal black hole
(\ref{5RN}) of Einstein-Maxwell theory.

By the way, the theory discussed above
is far from the only supergravity theory in ten
dimensions.  It is a particular kind called `type IIA.'  The `II' refers
to the fact that there are two independent gravitino fields.
In type IIA theory, these
gravitinos have opposite chirality.  This is turn allows type IIA
theory to be defined even on non-orientable manifolds.  There
is also a type IIB theory which has two gravitinos, but of the 
same chirality.  Thus, type IIB theory can only be defined on manifolds
with a global notion of chirality and, in particular, only on
orientable spacetimes.  We will discuss type IIB theory further in
section \ref{T} below.  Two other supergravity theories with less
supersymmetry are known as the type I and heterotic theories.  Each of these
types of supergravity in ten dimensions is associated with its own
version of string theory.  We will not discuss type I or heterotic
supergravity here, but a discussion of these theories and how they
are related to the type II theories can be found in \cite{Joe}.

\section{Branes in 9+1 type II Supergravity}
\label{10branes}
\label{branesols}

We now wish to discuss the basic brane solutions of type II supergravity
in 9+1 dimensions.  Since any solution of type IIA theory is really
a solution of eleven-dimensional supergravity (which just
happens to have a Killing field) in disguise, any brane solution of
type IIA theory immediately defines a brane solution of eleven
dimensional supergravity.  Thus, we should be able to construct the basic 
brane solutions of type IIA theory by working with the basic brane
solutions of section \ref{branes}.    For this reason we address
the type IIA solutions first in section \ref{IIAbranes}.
Next follows a short aside on brane singularities in section \ref{bsing}.
We then briefly discuss 
type IIB supergravity, and its relation through so-called
T-duality with the type IIA theory, in section \ref{T}.
When this is done, we will be in a position to provide a few
short comments in section \ref{pert} on perturbative string theory,
in order to give the traditional relativist a useful first
intuitive picture of the subject.  Below, we will discuss only
BPS branes, although there has recently been significant
interest in non-BPS D-branes\footnote{The term `non-BPS D-branes'
refers to a type of D-brane that have no BPS version, as opposed to just
the non-extremal branes obtained by adding energy to the BPS D-branes.}
\cite{non-BPS,non-BPSII}.

\subsection{The type IIA branes}
\label{IIAbranes}

It is useful here to recall our decomposition of the eleven-dimensional
metric and gauge field into the various fields of ten dimensional
supergravity.
We proceeded by projecting the fields along
various directions associated with
the Killing field.  In order to get
brane solutions of type IIA theory that are charged under all of the
type IIA gauge fields, a similar operation will need to be performed on
the brane solutions.  For any given brane in eleven dimensions, we
will need to reduce both a basic brane solution in which the Killing
field acts along the brane (i.e., is a symmetry of the brane), and one in
which it acts transverse to the basic brane.  

One may at first
wonder what it means for the brane to be transverse to the Killing field
since translations along a Killing field must leave the solution invariant,
and therefore must preserve the brane.  The answer to this puzzle
is the smearing mentioned in section \ref{smear}.  One can take a basic
brane solution, pick a direction transverse to the brane, and smear
the brane in that direction.  The result is an eleven dimensional
brane solution with a Killing field `transverse to the {\it basic}
brane.'  The Killing field means that the smeared brane is then
easily compactified and reduced to 9+1 dimensions.
All of the basic branes of type IIA supergravity are generated
by this procedure.  

Performing these reductions amounts to no more than using the
relations between the 9+1 fields and the 10+1
fields given in section \ref{KKred}
to write down the 9+1 solutions from the branes given
in section 
\ref{branes}.  We leave the details of the calculations to the reader, 
but we provide a list here of the various 9+1 brane solutions.  Below, 
we group together those branes charged under the Ramond-Ramond
gauge fields and those charged under the NS-NS gauge fields.  This 
grouping is natural from the
point of view of the type IIA theory (and of string perturbation theory), 
though we will see that it is somewhat less natural from the eleven
dimensional point of view.

Let us begin with the Ramond-Ramond branes.   It turns out that
type IIA theory has $p$-brane solutions with Ramond-Ramond charge
for every even $p$.  What is very nice is that, in terms of the
string metric, all of these solutions take much the same simple
form.   In order to treat all of the branes at once, it is useful
to introduce a uniform notation for both the electrically charged
branes and the magnetically charged branes.  For each
gauge field $A_n$, we can introduce (at least locally) a magnetic
dual gauge field $A_{9-n}$ through
$dA_{8-n} = \star F_{n+1}$.  A brane which couples magnetically to $A_n$
then couples electrically to $A_{9-n}$ and vice versa.  In type IIA
theory, this notation should introduce no confusion as the 
standard gauge fields have $n=1,2,3$ while these new (dual) gauge
fields have $n=5, 6, 7$.

Introducing the usual set of $p+1$ coordinates $x_\parallel$
along the brane and $9-p$ coordinates $x_\perp$ transverse to the brane
we have, for all even $p$, 
\begin{eqnarray}
\label{Dsol}
ds_{string}^2 &=& H_p^{-1/2} dx_{\parallel}^2 + H_p^{1/2} dx_{\perp}^2
\cr
A_{p+1} &=& - H_p^{-1} dx_{\parallel}^0 \wedge ... \wedge dx_{\parallel}^p \cr
e^{2 \phi} &=& H_p^{(3-p)/2}
\end{eqnarray}
where $H_p$ is a function only of the $x_\perp$ coordinates
and satisfies
\begin{equation}
\label{Dlap}
\partial_\perp^2 H_p =  - (7-p) \omega_{8-p} r_0^{7-p} \delta^{(9-p)}(x_\perp)
\end{equation}
for the basic brane solution.     Here $\omega_{8-p}$ is the volume
of the unit $(8-p)$-sphere and $r_0$ is the charge radius of the brane.
These are the solutions known as extreme R-R $p$-branes or, in a slight
abuse of language, as (extreme) D$p$-branes.  See section \ref{SandD} for an
explanation of this terminology.
As usual, we also obtain a
solution by considering any source term on the RHS of eq. (\ref{Dlap}).
For odd $p$, there are no gauge fields $A_{p+1}$ in type IIA supergravity
so (\ref{Dsol}) does not yield a solution to this theory for such 
cases\footnote{One might also ask about the case $p=8$, since we
have not discussed a 9-form gauge potential.  It turns out that
there is in fact a Ramond-Ramond 8-brane in type IIA theory and that
its existence is tied to the Chern-Simons term in the type IIA action.
In this work, we follow a policy of considering only the asymptotically
flat brane solutions, we will not discuss the 8-brane
explicitly.}.  Solutions for the non-extremal branes may be found in, 
e.g. \cite{YR}.  As  for the M-branes, their global structure is like 
that of the Schwarzschild solution as opposed to that of
non-extremal Reissner-N\"ordstrom.

Although all of these R-R branes take the same simple form (\ref{Dsol}),
they proceed by quite different routes from the 
eleven dimensional branes.  A short list follows:  The D0-brane
solution follows by reducing the smeared M-theory wave along the smearing
direction.
The D2-brane follows by reducing the smeared M2-brane along the smearing
direction.
The D4-brane is the reduction of the unsmeared M5-brane in a direction along
the brane. Finally, the D6-brane is the reduction of the unsmeared Kaluza-Klein
monopole along the $S^1$ fibers.

Next, there are the Neveu-Schwarz branes.  Since the only Neveu-Schwarz
gauge field is $A_2$, we expect to find two types of Neveu-Schwarz
branes.  The gauge field $A_2$ should couple electrically to a 1-brane
(a string) and it should couple magnetically to a 5-brane.  The
1-brane follows by reducing the M2-brane in a direction along the
brane.  The resulting solution
\begin{eqnarray}
\label{Fstring}
ds^2_{string} &=& - H_F^{-1} dx_\parallel^2 + dx_\perp^2 \cr
A_{2} &=& H_F^{-1} dx^0_\parallel \wedge dx^1_\parallel 
\cr
e^{2\phi} &=& H_F^{-1}
\end{eqnarray}
is known as the {\it fundamental string}.  The reason for this is
that this solution represents the classical limit of a long, straight
version of the {\it same} string that appears in string perturbation 
theory\footnote{The name `fundamental string' is, however, a bit
of a historical artifact as such strings are no longer regarded 
as significantly more fundamental than the other branes.  They are, 
however, quite useful due to the existence of string perturbation theory.}.

The Neveu-Schwarz 5-brane (NS5-brane) is constructed by smearing
the M5-brane in a transverse direction and then reducing along the 
smearing direction.  The result is
\begin{eqnarray}
\label{NS5}
ds^2_{string} &=& - dx_\parallel^2 + H_5 dx_\perp^2 \cr
F_{3} &=& - \frac{1}{3!} \partial_{x_\perp^i} H_5 \epsilon_{ijkl}
 dx^j_\perp \wedge dx^k_\perp \wedge dx^l_\perp 
\cr
e^{2\phi} &=& H_5.
\end{eqnarray}

An interesting property of the NS5-brane is that, in the string metric, 
the timelike Killing field has no horizon; its norm is constant across
the spacetime.  The would-be horizon at $x_\perp =0$ has receded
to infinite proper distance in all directions, not just along a Killing
slice as for the extreme Reissner-N\"ordstrom black hole.  As a result, 
the coordinate patch above actually covers a manifold that, in the
string frame, is geodesically complete\footnote{However, it not
geodesically complete either in the Einstein frame or as viewed from
the eleven-dimensional perspective.  In each of these cases, there
is a null singularity at the horizon.}.

Finally, there are the purely gravitational `branes' given by
the 9+1 versions of the Aichelburg-Sexl metric and of the Kaluza-Klein
monopole with all gauge fields set to zero and constant dilaton.  These
may be either written down directly by analogy with (\ref{Mwave})
and (\ref{11KK}) or constructed by
first smearing the 10+1 solutions along some $x_\perp$ direction and
reducing the result to 9+1 dimensions.    

This exhausts the possible
ways to make extremal 
9+1 branes by reducing (and perhaps smearing once) the basic
eleven-dimensional branes.    Below, we provide a few
words on their global structure and singularities.

\subsection{On brane singularities}
\label{bsing}

We have constructed the D-brane spacetimes from what, in many cases,
are nice eleven-dimensional solutions which are smooth except perhaps
at a singularity hidden inside a horizon.   However, the construction of
the reduced solutions introduces singularities.  Of the
9+1 branes, only the NS 5-brane does not have a naked
singularity\footnote{This statement refers to the metric in the string
frame.  In the Einstein frame there is a naked singularity on the horizon.
Its story is much like that of the D2-brane discussed below.}.
For the D4- and D6-branes and the fundamental string, 
this happens because the Killing field used in the reduction
has fixed points.  In the case of the D4 brane, we can see
from (\ref{M5}) that the norm of the
Killing field vanishes on the horizon of the M5-brane.  To some extent, this
simply means that the Killing field becomes null on the horizon, but a careful
analysis verifies that this Killing field does indeed have a fixed
point on the horizon.  Thus, when we identify points in the 10+1 spacetime 
under discrete Killing translations to get a spacetime with compact
Killing orbits, we create a ``Lorentzian conical singularity'' on the
horizon.  
As stated above, this is very much like the singularity that
arises on the horizon of the $M=0$ BTZ black hole when it is constructed
by identifying $AdS_3$ under translations along a Killing field.

A conical singularity 
is perhaps not so bad from the eleven dimensional perspective
but, because the norm of $\lambda^\mu$ vanishes at the fixed point, 
$e^\phi$ vanishes there as well.
The result is that, when expressed in terms of the string metric, 
the D4-brane has a (null) curvature singularity on its `horizon.'  The
resulting conformal diagram takes the form below.  The story of
the fundamental string is much the same.

\vbox{\centerline{\epsfbox{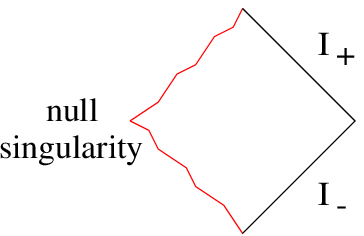}} \centerline{Fig 5. Conformal Diagram
for the extreme D4-brane or fundamental string.}}

\medskip

In the case of the D6-brane, the Killing field used in the reduction
of the Kaluza-Klein monopole clearly has a (6+1)-plane of fixed points
at $x_\perp=0$ even for the singly charged (smooth) case.    From the
geometry of the eleven-dimensional solution, it is clear that this
results in a 6+1 dimensional timelike singularity in the 9+1
dimensional D6-brane solution.  A short calculation shows that this
timelike singularity in fact resides at $x_\perp =0$ in 
(\ref{Dsol}).   One would be tempted to
call this location the `horizon' of the D6-brane as the norm of the
Killing field (as measured in the string metric) vanishes there, and
in fact this is the standard nomenclature.  However, one should remember
that for the D6-brane\footnote{And for the larger D-branes
which are not asymptotically flat and which we do not discuss here.}, this
horizon is a timelike
singularity.  The proper conformal diagram for the
D6-brane is therefore the one given in Fig. 6 below.

\vbox{
\centerline{\epsfbox{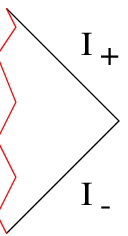}}
\centerline{Fig. 6. Conformal diagram for the extreme D6-brane.}}

\medskip

Let us now consider the D2-brane solution, which is the reduction of a
smeared M2-brane.
Although smearing the M2-brane in a transverse direction makes the
horizon of the eleven dimensional solution
 singular, one may take the perspective that the smeared solution
represents the approximate solution for an array of M2-branes
for which the associated charge radius is much larger than the
spacing between the branes.  In this case, one interprets
the D2-brane horizon as being non-singular\footnote{Actually, 
as consisting of many separate non-singular horizons.}  
from the eleven dimensional point of view.  However, 
from the 9+1 perspective, there is a null curvature on the horizon as
in the case of the D4-brane.  
Unlike the D4-brane, however, the curvature scalar $R$ of the string
metric does not diverge at the horizon and one must study more carefully
the null components of the curvature in order to see the singularity
explicitly.  The simplest way to detect the singularity from
the metric (\ref{Dsol}) is to note that the spheres around the brane
shrink to zero size at the horizon so that, if the solution were
smooth, the horizon could have only a single null generator, which is
impossible.  
Thus the conformal diagram of the 9+1 D2-brane
solution is again given by Fig. 5.

With an eleven dimensional perspective in mind, the singularities
of the D2-, D4-, and D6-brane solutions might not be considered
especially troubling.
Nevertheless, from the 9+1 perspective
the singularities are quite real and represent places where the
9+1 equations of motion break down.   Let us recall that, even
in eleven dimensions, the D0-brane or M-wave
solution {\it is} singular and should be thought of as describing
the approximate field produced by some `source.'  For the M-wave, 
one may think of this source as being a short wavelength
graviton, with the solution (\ref{Mwave}) itself representing just
the Coulomb part of the field.  Similarly, 
looking at the way that the D-brane
singularities interact with the equations of motion through
(\ref{Dlap}), it is natural to think of the singularities as
representing bits of matter, like a braney form of extremal dust, 
which are coupled to the supergravity.  One would then
consider the explicit solutions given above in (\ref{Dsol}) to
be degenerate cases in which the density of this dust of branes
has been taken to be distributional instead of smooth.  With a
smooth source, the solutions (\ref{Dsol}) yield smooth solutions
with a Killing field $\partial_t$ that is everywhere timelike, so that there
are no horizons.  

It is worth mentioning that, when the
dust distribution becomes too close to the delta-function of
(\ref{Dlap}), the curvature becomes very high near the center of this
distribution.  In the region where 
it is larger than the string (or Planck) scale, 
one no longer expects the classical supergravity metric to accurately
describe the physics.  However, when the dust configuration is close
to the delta-function case, the two supergravity solutions (smooth
and singular) will differ significantly only near the source and, in
particular, in the region where the curvature is beyond the string scale.
Thus, from a practical point of view, there is no
difference between considering the singular metrics (\ref{Dsol})
and smooth metrics given by smooth sources
which are close approximations to the delta-function.
The surprising thing from the perspective of familiar 3+1
Einstein-Maxwell theory is that a distributional dust
configuration makes sense and that it leads
to `point-like' objects instead of to black holes with a non-zero
horizon area.  This difference can be traced to the existence of the
dilaton field, which also diverges at the horizon.

By the way, this dual perspective of thinking of branes either
as solitonic objects intrinsic to some basic version of the theory
(like supergravity or string theory) or as external objects or
sources coupled to such a theory is pervasive in current
work on string/M-theory.  It is arguable that there
is no `right' choice to make and that in fact
some sort of self-consistent combination of the two perspectives
is more appropriate.  For the purposes of our exposition here, however, 
it is useful to keep the perspective that the branes
are solitonic objects in eleven dimensional supergravity\footnote{Except
perhaps for the M-wave, as discussed above.} but that, 
to properly represent them in ten dimensional Kaluza-Klein
reduced theories it is useful
to consider them to be extra bits of extremal `dust' that we couple
to the 9+1  supergravity theory. 

\subsection{T-duality and the type IIB theory}
\label{T}

The other type of maximally supersymmetric gravity theory in 9+1
dimensions is called type IIB theory.  It is not given by the dimensional
reduction of a 10+1 theory, though it has many of the same properties
as the type IIA theory.  For example, the subtheories of IIA and IIB
supergravity obtained by setting the Ramond-Ramond fields to zero are
identical.   One difference, however, is that
whereas type IIA supergravity
has Ramond-Ramond gauge fields of odd rank, the type
IIB theory has Ramond-Ramond gauge fields of even rank. 

The full IIA and IIB string theories are related by a symmetry
called T-duality.  At the level of supergravity, this is a symmetry
that maps solutions of type IIA theory with a Killing field into
solutions of type IIB theory with a Killing field.  It appears
that T-duality is an exact symmetry of the underlying string/M-theory
whether or not there is a Killing field but that, in the case
where there is no Killing field, 
a nice, nearly classical spacetime
is mapped by T-duality into a complicated highly quantum mechanical state.
It is only in the presence of a Killing symmetry that this duality maps
classical spacetimes to classical spacetimes.  T-duality is a true
duality in the sense that the transformation squares to the identity.

It is useful to first write down the explicit action of T-duality on
the metric and Neveu-Schwarz fields.  Let us introduce a coordinate $z$ 
such that translations in $z$ are a Killing symmetry. 
Let $x^\alpha$
be any other collection of coordinates which makes $(z,x^\alpha)$ a
coordinate patch.  Here we write the anti-symmetric Neveu-Schwarz
field as $B$ instead of $A_2$.  If the original solution is $(g,B)$,
then the transformed solution $(\tilde g, \tilde B)$ is given \cite{Bus} by
\begin{eqnarray}
\label{Tdual}
\tilde g_{zz} &=& 1/g_{zz}, \ \ \ \tilde g_{z\alpha} = B_{z \alpha}/g_{zz}, \cr
\tilde g_{\alpha \beta } &=& g_{\alpha \beta} - ( g_{z\alpha}
g_{z\beta} - B_{z\alpha}B_{z\alpha})/g_{zz},  \ \ \
\tilde B_{z\alpha} = g_{z \alpha}/g_{zz}, \cr
\tilde B_{\alpha \beta} &=& B_{\alpha \beta} -  (g_{z\alpha}B_{\beta z}
- g_{z\beta}B_{\alpha z})/g_{zz}, \ \  \tilde{\phi} = \phi + \log g_{xx}.
\end{eqnarray}
Note in particular that T-duality essentially interchanges the $g_{z \alpha}$
part of the metric with the $B_{z\alpha}$ part of the gauge field.  Now, in
the asymptotically flat context, the $g_{z0}$ component of the metric
is associated with momentum in the $z$-direction while the $B_{z0}$
component of the gauge field is associated with electrically charged
strings that extend in the $z$-direction.  Thus, one finds \cite{HGS} that
T-duality interchanges momentum, and the associated
Aichelburg-Sexl type solutions, with
fundamental strings (which carry the electric charge to which $B$ couples).

Strictly speaking, the T-duality of string theory
requires a Killing field with compact
($S^1$) orbits, through (\ref{Tdual}) maps solutions to solutions in
any case.  The original spacetime should be asymptotically flat
and, if the original $z$ coordinate is identified such that the length
of the $S^1$ at infinity is $L$, then the $z$ coordinate of the
transformed spacetime should be identified such that the length of the
$S^1$ at infinity is $\frac{4 \pi^2 l_s^2}{L}$.  The point is that,
if the orbits of the Killing field are compact, then
quantum mechanics implies that the momentum component around the compact
direction is quantized.  The proper normalization guarantees that 
T-duality  takes a solution with one quantum of momentum
to a solution containing a single fundamental string. 

Since the Neveu-Schwarz gauge fields are the same in
both the IIA and IIB theories, the branes carrying only Neveu-Schwarz charges
and the purely gravitational branes
also have identical forms.  On the other hand, the Ramond-Ramond
fields differ somewhat.  
Since the IIB theory has Ramond-Ramond gauge fields of even rank, it 
has R-R $p$-branes for odd $p$; i.e. 1-branes, 3-branes, 5-branes, 7-branes, and
9-branes.  It turns out that these solutions again take the form
(\ref{Dsol}), but now with odd $p$.  
The 3-brane is notable as it has constant dilaton and
a non-singular horizon.  In fact, the conformal diagram for the
3-brane looks essentially like Fig. 4, the conformal diagram for the
M5-brane.   For this reason, the D3-brane is closely associated with
5+1 anti-De Sitter space.

The effect of T-duality on the Ramond-Ramond fields is as follows:
\begin{eqnarray}
\label{RRT}
\tilde F_{n,\alpha_1 ...\alpha_n} = (const) F_{n+1, z \alpha_1 ...\alpha_n},
\cr
\tilde F_{n,z \alpha_1 ...\alpha_{n-1}} = (const) F_{n-1, \alpha_1 
...\alpha_{n-1}}.
\end{eqnarray}
Thus, if one takes a D$p$-brane and T-dualizes
in some direction along the brane, one obtains
a D$(p-1)$-brane solution which is smeared along the T-duality direction.
In string theory, D-brane charge is quantized. 
The normalization constants in (\ref{RRT}) are chosen so that,
and if the original solution
had one unit of D$p$-brane charge, then the transformed solution
has one unit of D$(p-1)$-brane charge.
Similarly, if one smears a unit charge
D$p$-brane solution in a transverse direction 
keeping the total charge equal to one quantum,
the T-dual solution is a unit charge D$(p+1)$-brane.  
See \cite{Joe} for a discussion of D-brane tensions, charge
quanta, etc.

\section{Some Remarks on D-brane Perturbation Theory}
\label{pert}

In the preceding sections we have discussed the supergravity aspects
of the various branes, including the D-branes. However, 
the real power of D-branes, and thus their importance, stems
from the fact that there is a renormalizable (in fact, order by order finite)
quantum perturbation theory to complement the classical supergravity
description.  This perturbation theory
describes both the internal dynamics of D-branes and their interactions with
the supergravity fields.   While not yet
a full quantum theory, it does allow one to perform certain interesting
calculations.  In particular, this is the key to the famous
counting of states of BPS and near-BPS black holes. 

Thus, although we will not discuss the details,
it is worthwhile to say a few words here about this perturbation
theory.  Below we will attempt to give, in language appropriate
to an audience of relativists, 
an intuitive understanding of how string perturbation theory is supposed to
be viewed.
We hope this gives a useful complement to standard presentations
which concentrate more on the perturbation theory details.

\subsection{Background field expansions and perturbative string theory}
\label{bfe}

The proper framework from which to view string perturbation theory
is that of the background field expansion (see e.g. \cite{Bryce}).  
Let us first review this idea in the context of standard quantum
field theory.  For definiteness, the reader may choose to focus on a
familiar low dimensional
interacting scalar field theory or even quantum mechanics.  
We will use $\phi$ to denote
the scalar field or, more generally, as a schematic notation for 
the collection of all relevant fields.   

Let us begin by supposing that there is some complete quantum
theory of this field, consisting of a set of field operators
$\hat \phi(x)$ and an associated set of composite operators acting on
a Hilbert space.  Exact calculations for interacting quantum
field theories are seldom possible, and one must resort to various
approximation schemes and expansions in small parameters in order
to obtain results.  For situations where the field is nearly in
its vacuum state, standard perturbation theory (see e.g. \cite{ItzZub})
can be a useful technique.

However, this is not the only case
of interest.   For example, it may be that a laboratory device (or a
star, black hole, or astrophysical event) produces a large, essentially
classical, disturbance in the field $\phi$ and that one wishes to
study small quantum effects in the resulting behavior.  It is in
such a regime that background field methods are useful.  
One first  considers the solution $\phi_0$ to the classical
field equations that would describe the situation if $\hbar$ were set
to zero.  One then rewrites the theory in terms of the field
$\widehat{\delta_0 \phi}(x) =
\hat \phi (x) - \phi_0(x)$.  Assuming that there is in fact a set
of semi-classical states in which the expectation value of 
$\hat \phi(x)$ is close to $\phi_0(x)$ and in which the fluctuations
are `small,' it makes sense to attempt a perturbative treatment in
terms of the field $\widehat{\delta_0 \phi}$.  

This is the
basic idea behind the background field expansion.  However, there is
one additional subtlety.  Although
one expects any differences to vanish
as $\hbar \rightarrow 0$, there need not be any
state in which the expectation value of $\hat \phi(x)$ is exactly
$\phi_0(x)$.  In a perturbative framework, one assumes
that the difference between the actual expectation value and the
classical solution $\phi_0$ can be expanded in powers of $\hbar$
and one proceeds to solve for it at each order of
perturbation theory.  It is useful to take the expectation value
calculated at order $n$, which we write as $\overline \phi_n$, to
be an effective `classical field' and to work at order $n$ with the
perturbation $\widehat{\delta_n \phi} = \phi(x) - \overline \phi_n.$

Within the range of validity of this perturbation theory,
one can (see \cite{Bryce}) 
expand about a general classical solution $\phi_0$ and 
obtain, at order $n$ in perturbation theory, an `effective action'
for the `effective classical background field' $\overline \phi$.   
The variations of the effective action with respect to the effective
field yield the classical equations of motion for $\phi$ corrected
by terms of up to order $n$ in $\hbar$ such that the solutions
of these equations yield the expectation value of $\hat \phi(x)$
(to order $n$) in a semi-classical state.

One could also attempt to follow the same general framework
but to expand around some arbitrary field $\phi_0$ which is not
a solution to the classical equations of motion.  In this case, 
the field $\widehat {\delta_0 \phi}(x)$ is not small and the perturbation
theory will not contain anything like a stable vacuum.  Because
the variation of the action does not vanish at the chosen
background, the action contains a term linear in 
$\widehat {\delta_0 \phi}(x)$ which acts as a source.  
This typically leads to 
various infrared divergences in the perturbation theory since, when integrated
over all time, this source will produce an infinite number of particles.
Thus if, for some
reason, someone had handed us not the full classical dynamics of the field 
but only the equations of the perturbation 
theory around an arbitrary background, the classical solutions of the
theory would still be recognizable. 

String perturbation theory
is in fact a version of background field theory in which the
`strings' correspond to excitations of the field $\widehat {
\delta_0 \phi} (x)$.  However, the logical order of the background
field framework is reversed or, perhaps more accurately, turned inside
out.  Instead of starting with a classical theory, quantizing,  and
performing the background field expansion, one instead {\it postulates}
the perturbative expansion\footnote{More
accurately, the S-matrix corresponding to such an expansion.}
about any background field
and then reconstructs the `classical' dynamics of the background field
in the manner discussed above
from the condition that the perturbation theory is well-defined.

This seemingly odd logical structure makes more sense when one
recalls that string theory is not, at present, a complete theory
based on any particular set of fundamental principles or axioms.
Rather, it is really an accidentally discovered set
of mathematical phenomena
which seem to hang together and which appear to have something to
do with quantum gravity, the unification of forces, and so on.
The way that string perturbation theory arose historically was
through interest in QCD and possible `strings' of gauge field flux
that would connect quarks in hadrons.  While studying such strings, it
was discovered that they defined a perturbation theory which
was finite order by order and which contained a spin two particle
which could be interpreted as a graviton.  Since finding a
perturbative treatment of quantum gravity, or even constructing a new
theory of gravity which could be treated perturbatively, had
been a question of interest for some time, string theory 
presented a solution to this technical problem:  
Simply take this accidentally discovered 
perturbation theory and use it to construct an associated theory of 
quantum (and classical) gravity.  In the case of string theory, 
the postulated perturbation theory was used to construct not
only the classical dynamics of the various fields, but also
to deduce the classical field content itself.  The rest, as they say,
is history.

Our story of supergravity discussed in the previous sections is relevant
here because the dynamics of string theory reduces in a certain
classical limit to classical supergravity. 
A few fine points are worth mentioning briefly.  The first
is that, when viewed as a background field theory of the sort discussed
above, classical string theory actually contains not just the
fields of classical supergravity, but an infinite tower of massive
fields as well.  The masses of these {\it classical} fields are, however, 
on the order of the string scale (and therefore considered to be large).  Thus, 
one expects there to be a large
regime in which these fields are not independently excited.  Instead, the
heavy (massive) fields
are `locked' to the values of the massless fields.  At the extreme
end of this regime, the massive fields are completely irrelevant.
However, as one pushes toward the boundaries of this regime, 
the massive fields may still have some effect on the dynamics. 
If one solves the classical equations of motion for the heavy fields, 
one finds that they are disturbed slightly by the massless fields and then
in turn provide small sources for the massless fields.  This analysis, 
known in the lingo of path integrals
as `integrating out' the massive fields, 
leads to additional effective interactions between the
massless fields.  Such interactions are
non-local on a scale set by the masses of the heavy fields; i.e., on the
scale of the string length.   When expanded in a power series, they
lead to a series of higher derivative terms in the action suppressed
by powers of the string scale.  These are the so-called 
$\alpha'$-corrections. The parameter $\alpha'$ is equal
to $l_s^2$ where $l_s$ is the string length.

In this way the string scale explicitly appears in the dynamics of
classical string theory.  Now, it is true that in the `real world' the string
length is likely to be within a few orders of magnitude of
the Planck scale.  In principle, however, 
the two scales are completely independent and should not be confused.  
The string scale controls the 
corrections to classical supergravity caused by the tower of massive
fields and the (9+1) Planck scale is the true quantum scale.
Their ratio defines the string coupling $g_s$.
The regime in which string perturbation theory is useful is $g_s \ll 1$, 
in which the string length is much greater than the 9+1 Planck length.

\subsection{Strings and D-branes}
\label{SandD}

In order to describe how D-branes fit into this picture, we should say just
a few more words about the relation of strings to supergravity.
As mentioned above, strings provide
rules for constructing the perturbation theory about a given 9+1 supergravity
background.   Roughly speaking, one replaces
the Feynman diagrams (related to particles) of familiar perturbation theory
with a new sort of diagram related to strings.  For details, the
reader should consult \cite{Joe} or \cite{GSW}.
For most of our purposes below, it will
suffice to think about the strings as classical objects.  

One can conceive of two basic types of strings.
The first are the so-called closed strings, which 
at any moment of time have the topology $S^1$.  Thus, they
resemble a classical rubber band.  It turns out that the closed
strings define a consistent perturbation theory in and of themselves, and
that it is this case that leads to the type II supergravities on which
we have focused.  Another version of the closed string leads
to heterotic supergravity, which has half as much supersymmetry
as the type II theories.

One might also consider so-called open strings which, 
at any instant of time, have the topology of an interval.  In order
for the dynamics of such strings to be well-defined, one must 
specify boundary conditions at the ends.    A natural
choice is to impose Neumann boundary conditions to describe
free ends.  Such strings are quite similar to classical rubber bands
that have been cut open.  It turns out
that this type of string does not yield a consistent perturbation theory by
itself, as two open strings can join together to produce a closed
string.  When open and closed strings are taken together, a consistent
perturbation theory does result.  This theory is associated with
type I supergravity, having half as much supersymmetry as the  type II
theories.

The other type of boundary condition that one can impose at the end
of a string is the Dirichlet boundary condition, requiring the end
of the string to remain fixed at some point in space.  One can also
consider a mixture of Dirichlet and Neumann boundary conditions, 
insisting that the end of the string remain attached to some submanifold
of spacetime, but otherwise leaving it free to roam around the
surface.  Surfaces associated with such Dirichlet boundary conditions
are known as Dirichlet submanifolds;
i.e. D-branes.
Again, for a consistent perturbation theory,  one must
consider closed strings in addition to these open strings.
Since we have singled out this submanifold as a special
place in the spacetime, this perturbation theory should not 
describe an expansion about empty space.  However, 
there remains the possibility
that it can describe an expansion about a background in which
certain sub-manifolds are picked out as special; i.e., near
a background which includes certain brane-like features.
Recall that, as a background field expansion, this perturbation
theory should tell us about all of the dynamics of the background, 
including any dynamics of the branes.

To make a long story short, it turns out that the Dirichlet submanifolds
are sources of the Ramond-Ramond gauge fields and of the gravitational
field.  That is, they carry both stress-energy and Ramond-Ramond
charge.  Thus, one might expect that they have something to
do with the branes discussed earlier that carry Ramond-Ramond charge.
In fact, this D-brane perturbation theory is supposed to give the
expansion about a background that includes such a
charged gravitating R-R brane in the asymptotic regime of small string
coupling $g_s$, which controls the strength of all interactions.
The perturbation theory describes both the dynamics
of the bulk fields (roughly speaking, 
through the closed strings) and of the brane
itself (roughly speaking, through the open strings).
The two parts are coupled and interact.

\subsection{On branes and perturbative expansions}

There is more to be said about the relation between perturbative
expansions and spacetime brane solutions, and this subsection contains
a quick overview of  the relevant features.  It is an important
aspect of the full story, and certainly a favorite part of D-brane lore.
However, it is arguably tangential and perhaps not of central
interest to a relativist.   As a result, on a first reading one might
wish to skip directly to section \ref{entropy} for a few comments on
the accounting of black hole microstates.  

For those readers who wish to 
more fully appreciate the perturbative D-brane construction, we now
go back a bit and clarify an issue in closed string perturbation theory.
This theory is supposed to describe the expansion around some
supergravity background.  However, it turns out that string perturbation
theory is only understood around backgrounds in which the Ramond-Ramond
gauge fields vanish.  Since one can see the Ramond-Ramond fields in
the string perturbation theory about backgrounds in which they happen
to vanish, and
since their classical dynamics is in fact determined from the other fields
by supersymmetry, the viewpoint is that the lack of string
perturbation theory about backgrounds with nontrivial Ramond-Ramond
fields is purely a technical one. In fact, in the last
year, some steps toward solving this problem have been taken 
\cite{RR1,RR2,RR3,RR4,RR5,RR6}.
Nevertheless, it means that one cannot use standard closed string
techniques to do perturbative calculations about backgrounds
containing branes with Ramond-Ramond charge.
D-brane techniques thus provide the
only tools to access this case.

In contrast, one can, at least in principle, use standard closed
string perturbation theory to study the behavior of excitations about
branes carrying only Neveu-Schwarz charge.  For electric NS charge (the
fundamental string) this works quite well while for magnetic
NS charge (the NS 5-brane) it turns out that such techniques cannot 
access the most interesting physics.

There are
several relevant points here.
One is that the effective coupling constant in string theory is not
really just the constant $g_s$, but is in fact $g_s e^{\phi}$;
i.e., it can vary over the classical solution.
Now, a perturbation theory is useful only at weak coupling, so
we imagine taking $g_s$ to be very small (with the asymptotic
value of the dilaton field set to one).   Perturbation theory about
the background can only be useful if the dilaton is such that the 
effective coupling remains small over the entire spacetime.
The other relevant point has to do with charge quantization in string/M-
theory, as we will discuss below.

Perturbation theory
about the fundamental string solution works quite well since, for 
small $g_s$, the effective coupling is small everywhere in the
spacetime.  In fact, the dilaton is such that the effective
coupling goes to zero near the singularity.   This will be true
for any value of the string charge radius.

Nevertheless, it is an interesting question to ask whether 
the inherent strength of the fundamental string solution should change
as we take $g_s$ to zero.  For the fundamental
string solution discussed in section \ref{10branes},
the charge radius is set by 
the strength of the delta-function in the source $\rho$.
As $\rho$ represents the quantity that stands on the right-hand side
of the (super) Einstein equations,
the coefficient $r_0^6$ of the delta-function 
may be thought of as being of the
form $GT$, where $G$ is the ten-dimensional gravitational constant and
$T$ is the tension (mass per unit length) of the fundamental string.

In string theory,
as in any theory with both
electric and magnetic charges, the charge is quantized in integer
multiples of some fundamental charge.  The BPS bound relates
the charge to the tension, so the tension of a BPS
object is quantized as well.  
For the fundamental string with $n$ units of charge, 
this tension is proportional to $n$ and does not depend on the string
coupling.  Recall that $G$ is proportional to $g_s^2$.
Thus, if we consider a string with a fixed number $n$ of charge units, 
the parameter $r_0$ that controls the departure of the supergravity
fields from flat empty space vanishes in the limit of weak string
coupling.  As a result, weak coupling perturbation theory in the fundamental
string background reduces, at least away from the singularity,
to just perturbation theory about flat space. As mentioned earlier, 
the singularity of the fundamental string solution is supposed to
represent a source due to the presence of fundamental strings.
Thus, perturbation theory about the fundamental string background is
naturally associated with string perturbation theory about flat space, 
in a sector where the strings are arranged to have
$n$ units of electric Neveu-Schwarz charge.

Now, the story for the NS 5-brane is quite different.  This time, 
the charge quantization specifies that, with $n$ units of charge, the
tension of the brane is proportional to $n/g_s^2$.  Thus, $GT \sim n$
and the supergravity fields remain unchanged as we take 
$g_s \rightarrow 0$.  Reading off the dilaton behavior from the
5-brane solution (\ref{NS5}), we see that the effective string coupling
becomes of order one at some finite place in the spacetime and 
then diverges as we look deep into the throat of the 5-brane.
As a result, string perturbation theory will not be a useful tool
to study dynamics associated with the throat of the NS 5-brane.

The Ramond-Ramond branes are a nice intermediate case between
these two.  It turns out that the tension of any R-R brane (with $n$
units of charge) is proportional to $n/g_s$.  Thus, $GT \sim n g_s$
goes to zero at weak coupling.  As a result, the supergravity
fields go over to flat empty space in this limit.  This means that, even
if at finite $g_s$ the dilaton forces the effective coupling to
diverge at the singularity of the brane, in the $g_s \rightarrow 0$ limit
the effective coupling is small over the entire spacetime (except at
the singularity itself).  On the other hand, since the mass per unit
volume of the D-brane is diverging, any internal dynamics associated
with motion of the D-brane
is frozen out in this limit; the actual dynamics at finite
$g_s$ may be thought of as perturbations around an infinitely massive
and therefore non-dynamical object.  The picture that one obtains
strongly resembles the D-brane picture described above; it consists
of flat empty space with a preferred submanifold in spacetime occupied
by a non-dynamical brane.  Thus, one might suppose that the D-branes
of perturbation theory should be identified in this way with the
Ramond-Ramond branes of supergravity.  Additional evidence for this
picture comes from the great success of D-brane perturbation theory
in reproducing the entropy of black 
holes \cite{SV,CM,HS2,BMPV,MS,JKM,BLMPSV,KT,JM,HM1,HM2}, hawking radiation
\cite{DMW,DM1,DM2}, and
the so-called grey-body factors \cite{MSII}
associated with the Ramond-Ramond branes.

\subsection{A few words on black hole entropy}
\label{entropy}

This is not the place for an in-depth discussion of just how D-brane
perturbation theory can be used to reproduce the properties of
supergravity solutions.  Such treatments can be found in \cite{JMPHD}
and in \cite{Joe}.  They involve the fact that the open strings
associated with D-branes describe, in the low energy limit, a certain
non-abelian Yang-Mills theory.  The low energy limit of that theory
can then be analyzed and used to study the low energy limit of the
brane dynamics.  Since BPS branes have the minimal possible energy
for their charge, this means that BPS and nearly BPS branes can
be addressed by such techniques.

We will, however, close by giving some parts of the entropy calculation for
a particular case.   As has already been mentioned, the solution
(\ref{3M2}) with three mutually orthogonal sets of M2-branes is
the simplest BPS black brane solution with non-zero entropy.  Let us compactify
a circle along one of the M2-branes (say, the one associated with the
$z_\parallel$ coordinates) and Kaluza-Klein reduce to 9+1 type IIA
supergravity.  Then, as we have seen, the $z$-type M2-branes (which are
wrapped around this circle)    
become fundamental strings in the IIA description while the 
$x$- and $y$-type M2-branes (which do not wrap around the compact circle)
become D2-branes.  It turns out that a simple description
of the microscopic perturbative states can be obtained by T-dualizing
this solution to the IIB theory along the direction in which the 
fundamental strings point.  This turns the fundamental strings into
momentum and the two sets of D2-branes into D3-branes.

Let us now
T-dualize twice more in, say, the two $y_\parallel$ directions.
This again yields a solution of IIB theory.  The momentum remains
momentum in the same direction, but one of the sets of D3-branes
has become a set of D1-branes and the other has become a set of D5-branes.
The D1-branes (D-strings) are stretched in the
same direction that the momentum is flowing, and this all happens
in one of the directions along the D5-branes.  These T-dualities
do not change the integer charges $Q_x,Q_y,Q_z$ associated with the various
types of branes: $Q_x$ is now the number of D5-branes, $Q_y$ the number
of D-strings, and $Q_z$ the number of momentum quanta.  One can check
that these T-dualities do not change the Bekenstein-Hawking entropy and, 
as supposed symmetries of the underlying string theory, they cannot 
change the number of microstates.

The case of a single D5-brane is particularly simple
to discuss.  It turns out that
the low energy dynamics
reduces to what is effectively just a collection of D-strings\footnote{Or,
even better, to a single D-string wrapped $Q_y$ times
around the direction in which the momentum flows. See, e.g.,
\cite{JMPHD}.} which
are stuck to the D5-brane but free to oscillate within it.  The momentum
in the solution is just the momentum carried by these oscillations, and
the energy of the solution is a linear sum of contributions from the
D5-brane rest energy, the D-string rest energy, and the momentum.
For a supersymmetric solution, all of the oscillations must move
in the same direction along the D-string and so are described by, say, 
right-moving fields on a 1+1 dimensional spacetime.
Oscillations of D-strings propagate at the speed of light and so the 
associated energy-momentum vector is null.

Thus, for
each string, one has $4$ massless $1+1$ rightmoving scalar fields corresponding
to the four internal directions of the fivebrane.  Supersymmetry
implies that there
are also four massless $1+1$ rightmoving
fermionic fields for each D-string.  A fermion
acts roughly like half of a boson, so we may think of this as $6Q_y$ massless
right-moving scalars
on $S^1 \times R$ (the worldvolume of a one-brane).   
A standard formula tells us that, given $n$ massless rightmoving scalars 
with $Q_z$ units of momentum, 
the entropy at large $Q_z$ 
is $S = 2\pi \sqrt{Q_z n/6}$.  Thus we have 
$S = 2\pi \sqrt{Q_yQ_z}$, in agreement
with the Bekenstein-Hawking entropy $S = A/4G_{11}$ (see
eq. (\ref{BHE}) with $Q_x =1$) for the associated black hole.

This gives an idea of the way in which D-brane perturbation theory provides
a microscopic accounting of the entropy of this BPS black hole.
The other BPS and near-BPS cases are similar in many respects.
It is quite satisfying to arrive at exactly the Bekenstein-Hawking
entropy formula without having to adjust any free parameters.
However, one is certainly struck by the qualitative differences
between the regime in which we are used to thinking about black holes and
the regime in which the string calculation is done.  We usually 
consider black holes with large smooth horizons.  In contrast, the perturbative
calculation is done in the asymptotic regime of small $g_s$, 
where spacetime is flat and the horizon has degenerated to zero size.
The belief is that supersymmetry guarantees the entropy of the quantum
system to be independent of $g_s$, as it does for other non-gravitational
systems\footnote{As supporting evidence, recall that the Bekenstein-Hawking
entropy of our BPS black hole does not depend on $g_s$ when written
in terms of the integer charges (\ref{BHE}).}.  
In any case, there is much room for speculation and investigation
in trying to match these pictures more closely and in understanding just
what form these states take in the black hole regime of finite $g_s$.

\vskip 1cm

{\centerline {\bf Acknowledgements}}

The author wishes to thank Gary Horowitz, Rob Myers,
Amanda Peet, and especially Joe
Polchinski for many discussions on string theory over the years.
Thanks also to both the participants of the Mazatl\'an winter school
and the ITP workshop on classical and quantum physics of strong
gravitational fields (Santa Barbara, 1999) for listening to various
versions of this material and for asking numerous questions.
Thanks especially to Abhay Ashtekar, Rafael Sorkin, and Robert Wald
for their encouragement to write up these notes.  
This work was supported in part by NSF grant PHY 9407194 to the ITP (Santa
Barbara), NSF grant PHY 9722362 to Syracuse University, and by funds
from Syracuse University.

\appendix

\section{Guide to Further Reading}
\label{read}

The following is a brief guide to a small fraction of the available literature
on strings, branes, and M-theory.  Other useful references will
undoubtedly be forthcoming as well.  In particular, the 1999
TASI summer school was devoted to ``Branes, Strings, and Gravity.''
The written versions of many of the lectures presented there should
appear on hep-th in the fall of 1999 (most likely with TASI in the title)
and should make useful reading.
I would guess that the talks by Clifford Johnson and Amanda Peet
might be most attuned to the interests and background of relativists, though
most of the lectures should probably be recommended as further reading
when they appear.

\bigskip

{\bf General:} The best and most recent  general reference available 
on string theory is Joe Polchinski's two-volume work ``String Theory''
\cite{Joe}. 
This book has an intuitive and pleasantly chatty style, but is
of course not aimed at an audience of relativists.  It contains much
discussion on spacetime aspects of string/M-theory, though
generally at a rather conceptual level and details of gravitating
solutions are not emphasized.  The book provides a useful
perspective on many modern topics, but predates the Maldacena conjecture
\cite{MC} (aka AdS/CFT).
A constructive way to read this book for a relativist with
a shortage of time might be to begin with volume II and to simply
take for granted all statements that arise from quantum field
theoretic or string perturbation arguments.  The original
text on the subject by Green, Schwarz, and Witten \cite{GSW} is
also a useful reference on many subjects.

\bigskip

{\bf Branes and spacetime solutions:}
A rather thorough, but not particularly recent, review of brane solutions
in string/M-theory can be found in the 1997 manuscript \cite{YR} by Youm. 
It contains numerous references  to the original works.

\bigskip

{\bf D-branes, Yang-Mills Theory, and Black Hole Entropy:}
As far as the mechanics of this subject is concerned, a canonical
reference is Juan Maldacena's Ph.D. thesis \cite{JMPHD}, although
this of course does not include the more modern perspective based on
the Maldacena conjecture.  For a lighter overview, 
Joe Polchinski's book \cite{Joe} also contains the essential points. 

\bigskip

{\bf The Maldacena Conjecture:} Although we have not had a chance
to explore it here, the Maldacena conjecture (aka the AdS/CFT correspondence)
is an important part of current work in string/M-theory.
There is now a quite thorough
review of the subject \cite{rev} by Aharony, Gubser, Maldacena, Ooguri, and Oz.
The address both the motivations behind the conjecture and the
evidence in support of it (up to the time of writing).  Of course, 
the subject may continue to develop rapidly, but this review will remain
an excellent starting point.

\section{Conventions}
\label{conv}

Since there are a number of different conventions for dealing with
anti-symmetric tensor (i.e., $n$-form) fields,  
it is worthwhile to spell out explicitly the ones used here.
In the text, $A_n$ denotes an $n$-form potential, and $F_{n+1}$ denotes
the corresponding $(n+1)$-form field strength.  We have 
$F_{n+1} = d A_{n}$, though some use is made of a modified ``field strength''
$\tilde F_{n+1}$ defined in the text.
The components $A_{\alpha_1 \alpha_2 \alpha_3 ...}$ of $A$ are
related to the $n$-form $A$ by

\begin{equation}
A_n = \frac{1}{p!} A_{\alpha_1 ...\alpha_n} dx^{\alpha_1} \wedge ... \wedge 
dx^{\alpha_n},
\end{equation}
so that we have
\begin{equation}
\int A_n = \int A_{0123...(n-1)} d^nx.
\end{equation}

It is also useful to define
\begin{equation}
|F_{m}|^2 = \frac{1}{m!}  g^{{\alpha_1} {\beta_1}} ...g^{{\alpha_1} {\beta_1}}
F_{\alpha_1 ....\alpha_m}F_{\beta_1 ....\beta_m}.
\end{equation}

The conventions used here are identical to those in
Polchinski's book \cite{Joe}.

\end{document}